\documentclass[review,3p,times]{elsarticle}
\usepackage{amsmath,hyperref}
\usepackage{lineno}
\usepackage{color}
\usepackage{subfigure}
\usepackage{epsfig}
\usepackage{caption,xcolor,float,graphicx}
\usepackage{graphics}
\usepackage{epstopdf}
\usepackage{float}
\usepackage{booktabs}
\usepackage{multirow}
\usepackage{bm}
\usepackage{amssymb}
\usepackage{booktabs}
\usepackage{dcolumn}
\usepackage{changepage}
\usepackage{chemarrow}
\usepackage{threeparttable}
\usepackage{comment}

\biboptions{sort&compress}

\usepackage[T1]{fontenc}
\journal{Applied Mathematics and Computation}

\begin{document}
\captionsetup[figure]{labelfont={bf},name={Fig.},labelsep=period}

\begin{frontmatter}
	
\title{Phase-field lattice Boltzmann method for two-phase electrohydrodynamic flows induced by Onsager-Wien effect}

\author[mymainaddress]{Mingzhen Zheng}
\author[mymainaddress]{Lei Wang\corref{mycorrespondingauthor}}
\cortext[mycorrespondingauthor]{Corresponding author}
\ead{wangleir1989@126.com}
\address[mymainaddress]{School of Mathematics and Physics, China University of Geosciences, Wuhan 430074, China}

\author[mymainaddress]{Fang Xiong}

\author[mysecondaddress]{Jiangxu Huang}
\address[mysecondaddress]{School of Mathematics and Statistics, Huazhong University of Science and Technology, Wuhan 430074, China}

\author[myfourthaddress]{Kang Luo}
\address[myfourthaddress]{School of Energy Science and Engineering, Harbin Institute of Technology, Harbin 150001, China}

\begin{abstract}
The leaky dielectric model is widely used in simulating two-phase electrohydrodynamic (EHD) flows. One critical issue with this classical model is the assumption of Ohmic conduction, which makes it inadequate for describing the newly discovered EHD flows caused by the Onsager-Wien effect [Ryu et al., Phys. Rev. Lett. 104, 104502 (2010)]. In this paper, we proposed a phase-field lattice Boltzmann (LB) method for two-phase electrohydrodynamic flows induced by the Onsager-Wien effect. In this scheme, two LB equations are employed to resolve the incompressible Navier-Stokes equations and the conservative Allen-Cahn equation, while another three LB equations are used for solving the charge conservation equations and the electric potential equation. After we validate the developed LB method, we perform a series of numerical simulations of droplet deformation under EHD conduction phenomena. Our numerical results indicate that the presence of the Onsager-Wien effect has a significant impact on droplet deformation and charge distribution. Also, it is interesting to note that, apart from the heterocharge layers near the electrodes, a charge cloud may form between the droplet interface and the electrode in some cases. To thoroughly understand the droplet dynamics, the effects of the reference length $d$, the applied voltage $\Delta\psi$, the permittivity ratio $\varepsilon_r$, and the ionic mobility ratio $\mu_r$ on droplet deformation and charge distribution are all investigated in detail.

\end{abstract}

\begin{keyword}	
Electrohydrodynamics; \; Two-phase flows; \; Onsager-Wien effect; \; Phase-field lattice Boltzmann method
\end{keyword}	 
\end{frontmatter}
	
\section{Introduction}\label{section1}
Electrohydrodynamics (EHD) is an interdisciplinary that studies the interaction between fluids and electric fields in a dielectric fluid medium \cite{D.A.Saville1997}. As a fundamental branch of EHD flows, multiphase EHD flows have been widely observed in many scientific and engineering domains, and play an important role in various applications such as solvent extraction \cite{K.P. Sci1992}, electrospraying \cite{AM JCR2021}, liquid fuel injection \cite{B.P.JCP2010}, inkjet printing \cite{P.V.R. ARFM 2014}, and biotechnology \cite{JGL BB2006}. In this context, understanding the response of a two-phase system  subjected to an electric field is of great significance for the optimal design and operation of EHD-based techniques used in these systems.

It is well established that EHD flows are generated by electric forces acting on charged particles within a liquid \cite{A.Castellanos1998}. Therefore, understanding how free charges are generated and transported in the system is the first step in uncovering the mechanisms behind multiphase flow motion in dielectric liquids. In the community of multiphase EHD flows, the most popular model used to describe the fluid movement generated by electric fields is the so-called leaky dielectric model \cite{Feng1999,Cui2019,Luo POF 2020,Liu2019}, which was first introduced by Melcher and Taylor \cite{Taylor 1969} under the assumption of Ohmic conduction \cite{Ryu2010}. The original leaky dielectric model combines the Stokes equations to characterize fluid motion with a current conservation equation that employs Ohmic conductivity \cite{D.A.Saville1997}. In this context, the charge accumulation at the multiphase interface occurs nearly instantaneously relative to the time scale of the fluid motion \cite{TomarJCP2007}. The force on these charges in the presence of electric field yields a net Maxwell stress within the interface that further drags fluids into motion \cite{D.A.Saville1997}. Since the introduction of the leaky dielectric model, numerous theoretical analyses and numerical studies have been carried out to investigate the multiphase EHD problems \cite{Feng1996,LH JCP2011,OS 2015}. The resulting data generally show qualitative agreement with experimental observations \cite{Feng1996,LH JCP2011,OS 2015}. However, extensive experiments have shown that the results obtained with the leaky dielectric model cannot give desirable quantitative comparisons \cite{D.A.Saville1997}. In addition to the effects of surface charge convection and electrokinetics explored in previous studies \cite{Feng1999,Cui2019,Luo POF 2020,Liu2019,Berry2013}, another possible reason is that the assumption of complete absence of volumetric charge in the original leaky dielectric model may not hold in certain cases \cite{D.A.Saville1997}. The experimental measurements of the current-voltage characteristics for dielectric liquids have shown that the Ohmic conduction regime holds only at relatively low voltages, where the current varies linearly with voltage \cite{FM 1991,Rubin PRL2008}. When the voltage continuously increases beyond the saturation voltage, the system enters the non-Ohmic conduction regime \cite{FM 1991,Rubin PRL2008}, where free charges not only appear at the multiphase interface but also exist in the bulk phase \cite{Luo PRE 2022}. In such a case, the free charges may be generated by either the injection mechanism or the conduction mechanism depending on operating regime of the system \cite{FM 1991,Rubin PRL2008}. In the case of the injection mechanism, free space charges are generated through complex electrochemical reactions occurring near the metal/liquid interface \cite{SJ2009}, with the resulting pumping effect commonly referred to as ion-drag pumping \cite{SIJeong 2003}. In contrast, for the conduction mechanism induced by the Onsager-Wien effect, free charges arise from the separation of ion pairs within the liquid, and the corresponding pumping is labeled as conduction pumping \cite{Atten2003,VazquezPOF2019}.

In recent years, many experimental and numerical investigations have been conducted to study the ion transport and EHD flow in single phase dielectric liquid, and the literature reveals that the majority of these studies focus on the charge injection induced electroconvection \cite{P.Atten 1996,Wu2013,Zhang2016,Luo PRE 2022}. However, charge injection-based ion-drag pumping can cause electrode and dielectric fluid deterioration, leading to non-uniform charge injection \cite{Feng 2004}. As a result, these EHD flows are deemed unreliable for long-term use \cite{Feng 2004}. Additionally, since the voltage required to activate the injection mechanism is typically much higher than that needed for the conduction mechanism, its potential safety concerns are another issue that needs attention \cite{Feng 2004}. Fortunately, EHD conduction pumping does not face the aforementioned issues, and presents distinct advantages over ion-drag pumping \cite{Feng 2004}. In light of this, some researchers have investigated the phenomenon of EHD conduction in dielectric liquids \cite{Atten2003,Feng 2004,Feng 2007,Fernandes 2014,VazquezPOF2019,WangAMM2021,Chen PRF 2023}. The first theoretical work on the phenomenon of EHD conduction may be attributed to Atten et al. \cite{Atten2003}, who found that there is a heterocharge layer with a finite thickness forms near the electrode. Later, Feng et al. \cite{Feng 2004} investigated the phenomenon of EHD conduction in dielectric liquids from both theoretical and experimental perspectives, and they showed that at high voltages, the impact of charge diffusion can be disregarded. Also, the authors performed an asymptotic theoretical analysis to assess the performance of the EHD pump under flow conditions, emphasizing the significant impact of fluid convection on charge distribution \cite{Feng 2007}. Subsequently, Fernandes et al. \cite{Fernandes 2014} investigated the EHD flow of dielectric liquids in the vicinity of a single cylindrical electrode, taking into account the Onsager effect in their model. More recently, V{\'a}zquez et al. \cite{VazquezPOF2019} discussed the underlying principles of EHD conduction pumps and proposed a new dimensionless parameter to characterize the operating state of the system. Wang et al. \cite{WangAMM2021} numerically studied the EHD conduction phenomenon in a square cavity equipped with a pair of asymmetry electrodes, and the influences of geometric scale and ionic mobility are investigated in detail. Following this work, Chen et al. \cite{Chen PRF 2023} further numerical investigated the EHD conduction phenomenon for viscoelastic fluids. 

Apart from the above mentioned single-phase EHD flows, recently few numerical works on multiphase EHD flows under EHD conduction phenomenon have also been reported \cite{Berry2013,Chirkov2021,Dobrovolskii2023}. Berry et al. \cite{Berry2013} proposed a hybrid numerical method to model the multiphase electrokinetic phenomenon without considering the Onsager-Wien effect. Chirkov et al. \cite{Chirkov2021} investigated the electrical deformation of conductive droplets suspended in low-conductivity liquid, and they showed that the conduction number affects the steady-state deformation of droplets. Building on this work, Dobrovolskii et al. \cite{Dobrovolskii2023} also studied how ion transport processes and differences in ionic mobility influence the behavior of water droplets in oil. While these limited studies provide meaningful insight into understanding the EHD conduction mechanisms, the interface tracking approaches appeared in the above works are confined to the traditional numerical methods, such as the finite element method \cite{Fernandes 2016,Chirkov2021,Dobrovolskii2023} and volume-of-fluid method \cite{Berry2013}, which usually require some extra efforts including interface reconstruction or reinitialization to capture the phase interface correctly \cite{AM2004}. As a mesoscopic numerical approach, the lattice Boltzmann (LB) method has been evolved into a promising tool for simulating complex flows, including multiphase flows \cite{X He 1999,Liang.PRE2014,LiangPRE2019} and porous media flows \cite{Chau PRE2006,liu CG2016,Ren ACS2022}. Compared with the traditional numerical approaches, the distinct advantages of LB method include easy of imposing boundary conditions and feasible for parallel programming \cite{SChen1998,Liang PRE2018}. In the literature, it is seen that the LB method has also been adopted by some researchers to investigate the multiphase flow dynamics in the presence of an electric field \cite{Cui2019,Luo PRE 2022}. Yet, we note that most of these available works are limited to the leaky dielectric model, in which the bulk free charges are neglected as stated previously \cite{D.A.Saville1997}. 

To the best of our knowledge, there is currently no established LB model that has been specifically applied to simulate two-phase EHD flows under the Onsager-Wien effect. To fill this gap, the primary goal of this paper is to propose an LB model for the two-phase EHD flows influenced by the Onsager-Wien effect. As demonstrated later, the numerical results obtained with the current LB model show good agreement with existing works. With the developed model, we further studied the droplet deformation by considering the Onsager-Wien effect. The remainder of this paper is structured as follows. The next section will describe the governing equations for two-phase EHD flows induced by the Onsager-Wien effect. Then, the dimensionless governing equations are introduced in  Sect. \ref{section3}. The details of the current LB model are given in Sect. \ref{section4}. Numerical tests to validate the present model are shown in Sect. \ref{section5}. Sect. \ref{section6} presents the main findings of our numerical results for droplet deformation under EHD conduction, and finally, we present a brief summary in Sect. \ref{section7}.

\section{Fundamental equations}

\label{section2}
In this section, we will briefly introduce the conservation equations for simulations of two-phase EHD flows induced by the Onsager-Wien effect. As a distinct multi-physical field coupling problem, the governing equations for two-phase EHD flows induced by the Onsager-Wien effect include the Navier-Stokes equations, the electric equations, the ionic transport equations, and the phase interface capturing equation.     

\subsection{Phase interface capturing equation}\label{section2.1}
The present work employs the phase-field based LB method to capture the phase interface, which is a popular approach in simulating two-phase flows in the LB community \cite{X He 1999,Liang.PRE2014}. Generally, the phase field method adopts the Cahn-Hilliard equation or the Allen-Cahn equation for interface capturing \cite{Chiu JCP 2011}. However, in contrast to the Allen-Cahn equation which only contains second-order derivatives, the fourth-order spatial derivative characteristic of the Cahn-Hilliard equation brings some inconveniences in numerical simulations \cite{AF2017}. In this setting, the recently proposed conservative Allen-Cahn equation given be Eq. (\ref{eq1}) is adopted in the present work. As pointed out by Chiu and Lin \cite{Chiu JCP 2011}, one of the main advantages of the conservative Allen-Cahn equation is that it can enforce the mass conservation \cite{Chiu JCP 2011}.
\begin{equation}
	\label{eq1}
	\frac {\partial \phi} {\partial t} +\nabla \cdot (\phi  \textbf{u} )=\nabla \cdot \left[M_{\phi}\left(\nabla \phi-\lambda \frac{\nabla\phi}{|\nabla\phi|}\right)\right],
\end{equation}
in which $\phi$ is the order parameter, and its value equals 0 and 1 in the bulk phase but varies smoothly across the interface. $\textbf{u}$ is the fluid velocity vector, $M_{\phi}$ is the mobility of the order parameter, $\lambda$ is a function of $\phi$ given by \cite{Liang PRE2018}
\begin{equation}
	\label{eq2}
	\lambda=\frac{4\phi(1-\phi)}{W},
\end{equation}
with $W$ representing the interface thickness. 

\subsection{Ionic transport equations}\label{section2.3}
For the EHD conduction considered here, induced by the Onsager-Wien effect, the Nernst-Planck equations are used to describe the charged species in EHD flows \cite{suh2012}
\begin{equation}
	\label{eq3}
	\frac{\partial c_+}{\partial t}+\nabla\cdot(\textbf{u}c_+ +\mu_+ c_+z_+\textbf{E}-D_+\nabla c_+)=R,	
\end{equation}
\begin{equation}
	\label{eq4}
	\frac{\partial c_-}{\partial t}+\nabla\cdot(\textbf{u}c_- +\mu_- c_-z_-\textbf{E}-D_-\nabla c_-)=R,
\end{equation}
in which $c_{+(-)}$ and $z_{+(-)}$ represent the concentration and valence of ions per unit volume, and the subscripts $-$ and $+$ are used to represent anions and cations, respectively. For the sake of simplicity in the analysis, we focus exclusively on monovalent ions, which implies that the valence of ions satisfies the condition $z_+=-z_-=1$ \cite{suh2012}. $\textbf{E}$ is the electric field. Additionally, $\mu_{+(-)}$ is the ionic mobility determined by $\mu_{+(-)} = e_0/[6\pi\eta R_H]$ \cite{A.Castellanos1998}, in which $e_0$ is the elementary charge, $R_H$ is the ionic equivalent hydrodynamic radius, $\eta=\rho\nu$ is the dynamic viscosity with $\nu$ and $\rho$ denoting the kinematic viscosity and the fluid density, respectively. $D_{+(-)}/\mu_{+(-)}=k_BT_r/e_0$ is the diffusivity \cite{A.Castellanos1998} (here, $k_B$ and $T_r$ are the Boltzmann constant andis the reference temperature, respectively). Further, $R$ is the reaction term whose formation is determined by considering the electric field-enhanced dissociation effect (also called Onsager-Wien effect in the literature) \cite{SIJeong 2003,VazquezPOF2019,WangAMM2021},  
\begin{equation}
	\label{eq5}
	R=\delta[c_0^2 F(E)-c_-c_+],
\end{equation}
with $\delta$ represents the compound constant calculated by $\delta=e_0(\mu_++\mu_-)/\varepsilon$, where $\varepsilon$ is the permittivity \cite{A.Castellanos1998,Y.K.Suh PRE2013}. $c_0=\sigma_0/(2e_0\overline{\mu})$ is the zero-field concentration with $\sigma_0$ and $\overline{\mu}=\sqrt{\mu_+\mu_-}$ \cite{suh2012} representing the zero-field conductivity and the average ionic mobility, respectively. $F(E)$ is the Onsager function, and its linear form can be expressed as \cite{Fernandes 2014,WangAMM2021}
\begin{equation}
	\label{eq6}
	F(E)=1+2\gamma E,
\end{equation}
in which $\gamma=e_0^3/(16\pi\varepsilon k_B^2T_r^2)$ is the Onsager constant and $E=|\textbf{E}|$ is the electric field strength \cite{Fernandes 2014,WangAMM2021}. 

\subsection{Governing equations for electric field}\label{section2.2}
Since the dielectric liquids are characterized by very low values of the conductivity, the dynamic current in the system is so small that the effect of magnetic induction can be ignored. In such a case, the electric field $\textbf{E}$ is irrotational ( i.e., $\nabla\times \textbf{E}=0$ ) \cite{A.Castellanos1998}, and the governing equations for electric field is given by \cite{A.Castellanos1998}
\begin{equation}
	\label{eq7}
\nabla\cdot\varepsilon\nabla\psi=-\rho_e,
\end{equation}
in which $\rho_e$ is the volumetric free charge density defined by \cite{A.Castellanos1998}
\begin{equation}
	\label{eq8}
	\rho_e=e_0(c_+-c_-),
\end{equation}
and $\psi$ is the electric potential, which is related to the electric field $\bf{E}$ as \cite{A.Castellanos1998}
\begin{equation}
	\label{eq9}
	\textbf{E}=-\nabla\psi.
\end{equation}
\subsection{Governing equations for incompressible fluid flow}\label{section2.4}
Assuming that two-phase fluid flows are immiscible and incompressible, the governing equations for the system can be formulated using the Navier-Stokes equations, which can be expressed as \cite{suh2012,A.Castellanos1998}
\begin{equation}
	\label{eq10}
	\nabla\cdot \textbf{u}=0, 	
\end{equation}
\begin{equation}
	\label{eq11}
	\frac {\partial (\rho \textbf{u})} {\partial t}+\nabla\cdot(\rho \textbf{uu})=-\nabla p+\nabla\cdot[\eta(\nabla \textbf{u}+\nabla \textbf{u}^T)]+\textbf{F}_s+\textbf{F}_e,
\end{equation}
in which $\textbf{u}$ is the fluid velocity, $p$ is the hydrodynamic pressure, and $\rho$ is the density of the fluid. In addition, $\textbf{F}_s$ is the surface tension force, which can be expressed as \cite{Kim JCP2005,Liang PRE2018}
\begin{equation}
	\label{eq12}
	\textbf{F}_s=\mu_\phi \nabla \phi,
\end{equation}
with $\mu _\phi$ being the chemical potential determined by \cite{Kim JCP2005,Liang PRE2018}
\begin{equation}
	\label{eq13}
	\mu_\phi=4\beta \phi(\phi-1)(\phi-0.5)-\kappa\nabla^2\phi,
\end{equation} 
Here, $\kappa$ and $\beta$ are two model parameters, which are related to the interface thickness $W$ and the surface tension $\zeta$ as $\kappa=3\zeta W/2$ and $\beta=12\zeta/W$ \cite{Kim JCP2005,Liang PRE2018}. Further, $\textbf{F}_e$ is the electric field force expressed as \cite{P.Atten 1996}
\begin{equation}
	\label{eq14}
	\textbf{F}_e=-\frac{1}{2}\textbf{E}^2\nabla\varepsilon+\rho_e\textbf{E},
\end{equation}
where the first term is the polarization force and the second term is the so-called Coulomb force \cite{P.Atten 1996,SIJeong 2003}.

\section{Dimensionless governing equations}\label{section3}
To simplify theoretical calculations and generalize the analysis results, we introduce the following dimensionless variables \cite{WangAMM2021}
\begin{equation}
	\label{eq15}
	\begin{split}
		\displaystyle	\textbf{x}&^*=\frac{\textbf{x}}{d},\quad \textbf{u}^*=\frac{\textbf{u}d}{\overline{\mu}\Delta\psi},\quad t^*=\frac{\overline{\mu}\Delta\psi t}{d^2},\quad p^*=\frac{d^2p}{\overline{\mu}^2\Delta\psi^2\rho},\quad M_\phi^*=\frac{M_\phi }{\overline{\mu}\Delta\psi},\\
		\displaystyle &\textbf{E}^*=\frac{d\textbf{E}}{\Delta \psi},\quad c_{+(-)}^*=\frac{c_{+(-)}}{c_0},\quad \rho_e^*=\frac{\rho_e}{e_0c_0},\quad \psi^*=\frac{\psi}{\Delta\psi},\quad \phi^*=\frac{\phi}{\Delta\phi},
	\end{split}
\end{equation}
where $d$ is the characteristic length of the system, and the variables with asterisk $*$ denote dimensional variables, and then the corresponding dimensionless governing equations can be written as
\begin{equation}
	\label{eq16}
	\frac{\partial\phi^*}{\partial t^*}+\nabla^*\left(\phi^*\textbf{u}^*\right)=\nabla^*\left[M_{\phi}^*\left(\nabla^*\phi^*-\frac{4\phi^*(1-\phi^*)\nabla^*\phi^*}{Cn|\nabla^*\phi^*|}\right)\right],
\end{equation}
\begin{equation}
	\label{eq17}
	\frac{\partial c_+^*}{\partial t^*}+\nabla^*\cdot\left[c_+^*(\textbf{u}^* +m_+\textbf{E}^*)\right] -\alpha_+\nabla^{*^2}c_+^*=2W_0\left[(1+2O\textbf{E}^*)-c_+^*c_-^*\right],
\end{equation}
\begin{equation}
	\label{eq18}
	\frac{\partial c_-^*}{\partial t^*}+\nabla^*\cdot\left[c_-^*(\textbf{u}^* -m_-\textbf{E}^*)\right] -\alpha_-\nabla^{*^2}c_-^*=2W_0\left[(1+2O\textbf{E}^*)-c_+^*c_-^*\right],
\end{equation}
\begin{equation}
	\label{eq19}
	\textbf{E}^*=-\nabla^*\psi^*,
\end{equation}
\begin{equation}
	\label{eq20}
	\nabla^{*^2}\psi^*=-W_0\rho_e^*,
\end{equation}
\begin{equation}
	\label{eq21}
	\nabla^*\cdot \textbf{u}^*=0,	
\end{equation} 
\begin{equation}
	\label{eq22}
	\frac{\partial \mathbf{u}^*}{\partial t^*} + \mathbf{u}^* \cdot \nabla^* \mathbf{u}^* = -\nabla^* p^* + \frac{Me^2}{T} \nabla^{*^2}\textbf{u}^*+W_0Me^2\textbf{E}^*(c_+^*-c_-^*)-\frac{1}{2}Me^2\nabla^*\textbf{E}^{*^2}+\frac{Me^2\nabla^*\phi^{*^2}}{4Ca_E}\left[\frac{48}{Cn}\left(\phi^*-1\right)\left(\phi^*-0.5\right)-\frac{3Cn\nabla^{*^2}}{2}\right].
\end{equation}
In this context, one can clearly find that the two-phase EHD flows induced by the Onsager-Wien effect are described by the following dimensionless parameters,
\begin{equation}
	\label{eq23}
	\alpha_{+(-)}=\frac{D_{+(-)}}{\overline{\mu}\Delta \psi}, \, Cn=\frac{W}{d},\, W_0=\frac{e_0c_0d^2}{\varepsilon\Delta\psi},\,
	Me=\frac{1}{\overline{\mu}}\left(\frac{\varepsilon}{\rho}\right)^{1/2},\, T=\frac{\varepsilon \Delta \psi}{\eta \overline{\mu}},\, m=\frac{\mu_-}{\mu_+},\, Ca_E=\frac{\varepsilon\Delta\psi^2r}{\zeta d^2},\, O=\frac{\Delta\psi\gamma}{d},
\end{equation}
In this context, $\alpha_{+(-)}$ denotes the dimensionless charge diffusion number for cations or anions, $Cn$ refers to the Cahn number. $W_0$ represents the conduction number, which is a crucial parameter that differentiates the two limiting states in EHD conduction (i.e., saturation state and ohmic state) \cite{VazquezPOF2019, WangAMM2021}. $Me$ stands for the dimensionless mobility parameter, which is solely determined by the intrinsic physical properties of the dielectric liquid \cite{WangAMM2021}. $T$ characterizes the electric Rayleigh number, which is the ratio of the viscous force and the Coulomb force \cite{WangAMM2021}. Additionally, $m$ is the ratio of $\mu_-$ to $\mu_+$. $Ca_E$ is the electric capillary number, which is the ratio of the surface tension force to the electric force \cite{Luo POF 2020}. Furthermore, $O$ is the non-dimensional Onsager number \cite{Y.K.Suh PRE2013}, which is influenced by the applied voltage. This voltage-dependent variation can inconvenience the process of accurately determining the operational state of the pump \cite{VazquezPOF2019}. To this end, V{\'a}zquez et al. \cite{VazquezPOF2019} proposed a voltage independent dimensionless number $\beta$ in the EHD conduction problem induced by the Onsager-Wien effect, which is given by
\begin{equation}
	\label{eq24}
	\beta^2=OW_0 \implies \beta=(OW_0)^{1/2}=\left(\frac{e_0^3\sigma_0d}{32\pi\varepsilon^2\overline{\mu}T_r^2k_B^2}\right)^{1/2}.
\end{equation}

\section{Lattice Boltzmann method}\label{section4}

In this section, the LB method for simulating two-phase EHD flows induced by the Onsager-Wien effect is presented. In terms of the collision operator in the LB community, the most popular LB model is the so-called single-relaxation-time (SRT) LB model \cite{LiangPRE2019,Chau PRE2006,liu CG2016,Ren ACS2022}. However, owing to the diffusive coefficient appearing in the charge conservation equation being significantly small, employing the SRT collision operator may fail to produce confident numerical results \cite{IK 2011,WangAMM2021}. Therefore, in this article, we employ the multiple-relaxation-time (MRT) collision operator, which has been proven to have better numerical stability and accuracy than the SRT model \cite{LS PRE2011}. 

In general, the LB model is performed in a regular Cartesian mesh with the D$n$Q$b$ ($n$-dimensional $b$-velocity) lattice. For the two-dimensional case considered here, the D2Q9 and the D2Q5 lattice models are employed, which are respectively defined as \cite{SChen1998,Zhan2021}

D2Q9:
\begin{equation}
	\label{eq25}
	\textbf{c}=c
	\begin{bmatrix}
		0 &1 &0  &-1 &0  &1 &-1 &-1 &1\\
		0 &0 &1  &0  &-1 &1 &1  &-1 &-1
	\end{bmatrix}
	,\,\omega_i=
	\left\{
		\begin{array}{cc}
			4/9, & i=0\\
			1/9, & i=1,\cdots,4\\
		    1/36,& i=5,\cdots,8
		\end{array}
 \right.
	, \, c_s=\frac{c}{\sqrt3},\\
\end{equation}\par
D2Q5:
\begin{equation}
	\label{eq26}
	\textbf{c}=c
	\begin{bmatrix}
		0 &1 &0  &-1 &0\\
		0 &0 &1  &0  &-1
	\end{bmatrix}
	,\, \overline{\omega}_i=
 \left\{
	\begin{array}{cc}
		1/3, & i=0\\
		1/6, & i=1,\cdots,4\\
	\end{array}
 \right.
	, \, c_s=\frac{c}{\sqrt3},
\end{equation}
with $c=\Delta x/\Delta t$ represents the discrete velocity, where $\Delta x$ is the lattice spacing step and $\Delta t$ represent the lattice time step. $\omega_i$ and $\overline{\omega}_i$ are the weight coefficients in the $i$-th direction, and $c_s$ is the sound speed of the lattice \cite{wang2022,Zhan2021}.

\subsection{Lattice Boltzmann method for phase interface capturing equation}
\label{section4.1}
To solve the conservative Allen-Cahn equation within the LB framework, we adopt the model proposed by Liang et al. \cite{Liang PRE2018} for its good numerical performance. Note that the LB method in Ref. \cite{Liang PRE2018} is constructed with the SRT collision operator, and the corresponding MRT-LB equation can be expressed as
\begin{equation}
	\label{eq27}
	h_i(\textbf{x}+\textbf{c}_i\Delta t, t+\Delta t)=h_i(\textbf{x},t)-\textbf{M}^{-1}\textbf{S}^h\textbf{M}\left[h_i(\textbf{x},t)-h_i^{eq}(\textbf{x},t)\right]+\Delta t\left[\textbf{M}^{-1}\left(\textbf{I}-\frac{\textbf{S}^h}{2}\right)\textbf{M}\right]H_i(\textbf{x},t),
\end{equation}
in which $h_i(\textbf{x},t)$ is the distribution function of the order parameter $\phi$ at position $\bf{x}$ and time $t$. $\textbf{I}$ is the identity matrix. $\textbf{M}$ is the transformation matrix derived from the discrete velocities via the Gram-Schmidt orthogonalization \cite{Chai2023}. For the D2Q9 lattice, the transformation matrix related to the natural moments can be chosen as \cite{Chai2023}
\begin{equation}
	\textbf{M}=
	\begin{bmatrix}
		1& 1& 1& 1& 1& 1& 1& 1& 1 \\
		0& 1& 0&-1& 0& 1&-1&-1& 1 \\
		0& 0& 1& 0&-1& 1& 1&-1&-1 \\
		0& 1& 0& 1& 0& 1& 1& 1& 1 \\
		0& 0& 1& 0& 1& 1& 1& 1& 1 \\
		0& 0& 0& 0& 0& 1&-1& 1&-1 \\
		0& 0& 0& 0& 0& 1&-1&-1& 1 \\
		0& 0& 0& 0& 0& 1& 1&-1&-1 \\
		0& 0& 0& 0& 0& 1& 1& 1& 1 
	\end{bmatrix}.
\label{eq28}
\end{equation}
$\textbf{S}^h$ is the diagonal relaxation matrix given by
\begin{equation}
	\label{eq29}
 	\textbf{S}^h=\text{diag}(s_0^h,s_1^h,s_2^h,s_3^h,s_4^h,s_5^h,s_6^h,s_7^h,s_8^h),
\end{equation}
where $s_1^h$ and $s_2^h$ are two parameters related to the mobility of the order parameter \cite{Chai2023},
\begin{equation}
	\label{eq30}
	\tau_h=0.5+\frac{M_{\phi}}{c_s^2\Delta t}=\frac{1}{s_1^h}=\frac{1}{s_2^h},
\end{equation}
and the other parameters in $\textbf{S}^h$ are free. In addition, $h_i^{eq}(\textbf{x},t)$ is the equilibrium function written as \cite{Liang PRE2018} 
\begin{equation}
	\label{eq31}
	h_i^{eq}(\textbf{x},t)=\omega_i\phi(1+\frac{\textbf{c}_i\cdot \textbf{u}}{c_s^2}).
\end{equation}
$H_i(\textbf{x},t)$ is the discrete forcing term given by \cite{Liang PRE2018}
\begin{equation}
	\label{eq32}
	H_i(\textbf{x},t)=\frac{\omega_i\textbf{c}_i\left[\partial_t(\phi\textbf{u})+c_s^2\dfrac{\nabla\phi}{|\nabla\phi|}\cdot\dfrac{4\phi(1-\phi)}{W}\right]}{c_s^2}.
\end{equation}
The order parameter $\phi$ is computed by \cite{Chen PRF 2023}
\begin{equation}
	\label{eq33}
	\phi=\sum_{i}h_i.
\end{equation}

\subsection{Lattice Boltzmann method for ionic transport equations}\label{section4.3}
Inspired by our previous works \cite{WangAMM2021}, the evolution equation of the LB method with MRT collision operator for ionic transport equations can be written as 
\begin{equation}
	\label{eq34}
	\begin{aligned}
	p_i(\textbf{x}+\textbf{c}_i\Delta t, t+\Delta t)=p_i(\textbf{x},t)-\textbf{M}^{-1}\textbf{S}^p\textbf{M}&\left[p_i(\textbf{x},t)-p_i^{eq}(\textbf{x},t)\right]+\\
	\Delta t&\left[\textbf{M}^{-1}\left(\textbf{I}-\frac{\textbf{S}^p}{2}\right)\textbf{M}\right]\left[R_i(\textbf{x},t)+L_i(\textbf{x},t)\right]-\textbf{M}^{-1}\textbf{S}^p\textbf{M} G_i(\textbf{x},t),
	\end{aligned}
\end{equation}
\begin{equation}
	\label{eq35}
	\begin{aligned}
	n_i(\textbf{x}+\textbf{c}_i\Delta t, t+\Delta t)=n_i(\textbf{x},t)-\textbf{M}^{-1}\textbf{S}^n\textbf{M}&\left[n_i(\textbf{x},t)-n_i^{eq}(\textbf{x},t)\right]+\\
			\Delta t&\left[\textbf{M}^{-1}\left(\textbf{I}-\frac{\textbf{S}^n}{2}\right)\textbf{M}\right]\left[\overline{R}_i(\textbf{x},t)+\overline{L}_i(\textbf{x},t)\right]+\textbf{M}^{-1}\textbf{S}^n\textbf{M}\overline{G}_i(\textbf{x},t),
	\end{aligned}
\end{equation}
in which $n_i(\textbf{x},t)$ and $p_i(\textbf{x},t)$ represent the distribution functions for anions and cations, respectively. The terms $\textbf{S}^p$ and $\textbf{S}^n$ correspond to the diagonal relaxation matrix given by \cite{Chai2023}
\begin{equation}
	\label{eq36}
	\textbf{S}^p=\text{diag}(s_0^p,s_1^p,s_2^p,s_3^p,s_4^p,s_5^p,s_6^p,s_7^p,s_8^p),\quad
	\textbf{S}^n=\text{diag}(s_0^n,s_1^n,s_2^n,s_3^n,s_4^n,s_5^n,s_6^n,s_7^n,s_8^n),
\end{equation}
where $s_1^{p,\;n}$, $s_2^{p,\;n}$ are the relaxation parameters determined by the diffusivity as \cite{Chai2023}
\begin{equation}
	\label{eq37}
	\tau_p=0.5+\frac{D_+}{c_s^2\Delta t}=\frac{1}{s_1^p}=\frac{1}{s_2^p},\quad \tau_n=0.5+\frac{D_-}{c_s^2\Delta t}=\frac{1}{s_1^n}=\frac{1}{s_2^n},
\end{equation}
and the other parameters in $\textbf{S}^{p,\;n}$ are free. In addition,  $p_i^{eq}(\textbf{x},t)$ and $n_i^{eq}(\textbf{x},t)$ are the equilibrium functions defined as \cite{WangAMM2021}
\begin{equation}
	\label{eq38}
	p_i^{eq}(\textbf{x},t)=\omega_ic_+\left[1+\frac{\textbf{c}_i\cdot\textbf{u}}{c_s^2}\right],\quad
	n_i^{eq}(\textbf{x},t)=\omega_ic_-\left[1+\frac{\textbf{c}_i\cdot\textbf{u}}{c_s^2}\right].
\end{equation}
Further, $R_i$, $\overline{R}_i$, $L_i$, $\overline{L}_i$, $G_i$ and $\overline{G}_i$ are the forcing terms given by \cite{WangAMM2021}
\begin{equation}
	\label{eq39}
	R_i(\textbf{x},t)=\overline{R}_i(\textbf{x},t)=\omega_iR,\quad
	L_i(\textbf{x},t)=\frac{\omega_i\textbf{c}_i\cdot c_+\textbf{F}}{c_s^2},\quad
	\overline{L}_i(\textbf{x},t)=\frac{\omega_i\textbf{c}_i\cdot c_-\textbf{F}}{c_s^2},
\end{equation}
\begin{equation}
	\label{eq40}
	G_i(\textbf{x},t)=\omega_i\textbf{c}_i\cdot\frac{c_s^2\ell_+}{c_s^2} \nabla\psi,\quad
	\overline{G}_i(\textbf{x},t)=\omega_i\textbf{c}_i\cdot \frac{\mu_-c_-}{c_s^2}\nabla\psi,
\end{equation}
with $\textbf{F}=\textbf{F}_s+\textbf{F}_e$ being total force. Finally, the concentrations of positive and negative ions can be obtained by \cite{WangAMM2021}
\begin{equation}
	\label{eq41}
	c_+=\sum_{i}p_i+\frac{\Delta t}{2}R,\quad 	c_-=\sum_{i}n_i+\frac{\Delta t}{2}R.
\end{equation}
As pointed out by previous work \cite{WangAMM2021}, since the reaction term $R$ is related to the charge density (i.e., Eq. (\ref{eq5})),  theoretically additional iterative steps are needed to determine the convergent solution. Yet, the quadratic nature of the above equations makes it possible to explicitly obtain the concentrations as \cite{WangAMM2021} 
\begin{equation}
	\label{eq42}
	c_+=\frac{(\lambda_0\overline{c}_+-\lambda_0\overline{c}_--1)+\sqrt{(1-\lambda_0\overline{c}_++\lambda_0\overline{c}_-)^2+4\lambda_0\overline{c}_+}}{2\lambda_0},\quad c_-=c_+-\overline{c}_++\overline{c}_-,
\end{equation}
in which $\lambda_0=0.5\Delta t\delta$ is a model parameter, $\overline{c}_+$ and $\overline{c}_-$ are two temporal charge densities computed by\cite{WangAMM2021}
\begin{equation}
	\label{eq43}
	\overline{c}_+=\sum_{i}p_i+\frac{\Delta t}{2}\delta c_0^2F(E),\quad \overline{c}_-=\sum_{i}n_i+\frac{\Delta t}{2}\delta c_0^2F(E).
\end{equation}

\subsection{Lattice Boltzmann method for electric field}\label{section4.2}
Following the work of Chai et al. \cite{Chai2008}, the evolution equation of the MRT-LB method for the electric field can be expressed as 
\begin{equation}
	\label{eq44}
	g_i(\textbf{x}+\textbf{c}_i\Delta t, t+\Delta t)=g_i(\textbf{x},t)-\textbf{N}^{-1}\overline{\textbf{S}}^g\textbf{N}\left[g_i(\textbf{x},t)-g_i^{eq}(\textbf{x},t)\right]+\frac{\Delta t\overline{\omega}_i\rho_e\vartheta}{\varepsilon}, 
\end{equation}
where $g_i(\textbf{x},t)$ represents the distribution function, $g_i^{eq}(\textbf{x},t)$ denotes the equilibrium function defined by \cite{Chai2008}
\begin{equation}
	\label{eq45}
	g_i^{eq}(\textbf{x},t)=
	\left\{
	\begin{array}{cc}
		(\overline{\omega}_0-1.0)\psi(\textbf{x},t), & i=0\\
		\overline{\omega}_i\psi(\textbf{x},t), &i\neq 0
	\end{array},
	\right.
\end{equation}
For the electric field, since the D2Q5 lattice model is employed, the transformation matrix  $\textbf{N}$ can be chosen as \cite{Chai2023}
\begin{equation}
	\label{eq46}
	\textbf{N}=
	\begin{bmatrix}
		1& 1& 1& 1& 1 \\
		0& 1& 0&-1& 0 \\
		0& 0& 1& 0& -1 \\
		0& 1& 0& 1& 0 \\
		0& 0& 1& 0& 1 
	\end{bmatrix}.
\end{equation}
In addition, $ \overline{\textbf{S}}^g=\text{diag}(s_0^g,s_1^g,s_2^g,s_3^g,s_4^g) $ is the diagonal relaxation martix with  $s_1^g$ and $s_2^g$ given by \cite{Chai2023}
 \begin{equation}
 	\label{eq47}
   	\tau_g=0.5+\frac{\vartheta}{c_s^2\Delta t}=\frac{1}{s_1^g}=\frac{1}{s_2^g},
 \end{equation}
where $\vartheta$ is the artificial diffusion coefficient, set to 0.3 to balance the evolution speed and stability in our simulations \cite{WangAMM2021}, while the other parameters in $\overline{\textbf{S}}^g$ are free. Further, the electric potential $\psi(\textbf{x},t)$ can be computed by \cite{Chai2008}
\begin{equation}
	\label{eq49}
	\psi(\textbf{x},t)=\displaystyle\sum_{i\neq 0}\frac{1}{1-\overline{\omega}_0}g_i(\textbf{x},t).
\end{equation}
Because of the inherent mesoscopic framework of the LB method, the electric field $\textbf{E}$ can also be computed locally \cite{Chai2008}
\begin{equation}
	\label{eq50}
	\textbf{E}=-\nabla\psi=\frac{\sum_i\textbf{c}_ig_i}{c_s^2\tau_g\Delta t}.
\end{equation}

\subsection{Lattice Boltzmann equation for incompressible fluid flow}\label{section4.4}
Inspired by previous work, the MRT-LB equation, which is employed to model incompressible fluid flow, can be expressed as \cite{Guo2002}
\begin{equation}
\label{eq51}
	f_i(\textbf{x}+\textbf{c}_i\Delta t, t+\Delta t)=f_i(\textbf{x},t)-\textbf{M}^{-1}\textbf{S}^f\textbf{M}\left[f_i(\textbf{x},t)-f_i^{eq}(\textbf{x},t)\right]+\Delta t\left[\textbf{M}^{-1}\left(\textbf{I}-\frac{\textbf{S}^f}{2}\right)\textbf{M}\right]F_i(\textbf{x},t),
\end{equation}
where $f_i^{eq}(\textbf{x},t)$ represents the equilibrium function defined by \cite{Guo2002}
\begin{equation}
	\label{eq52}
	f_i^{eq}(\textbf{x},t)=
	\left\{
	\begin{array}{cc}
		(\omega_0-1.0)\frac{p}{c_s^2}+\rho r_i(\textbf{u}), &i=0,\\
		\omega_i\frac{p}{c_s^2}+\rho r_i(\textbf{u}), &i\neq0,
	\end{array}
 	\right.
\end{equation}
with
\begin{equation}
	\label{eq53}
	r_i(\textbf{u})=\omega_i\left[\frac{\textbf{c}_i\cdot\textbf{u}}{c_s^2}+\frac{(\textbf{c}_i\cdot \textbf{u})^2}{2c_s^4}-\frac{\textbf{u}\cdot\textbf{u}}{2c_s^2}\right].
\end{equation}
$F_i(\textbf{x},t)$ denotes the discrete forcing term given by \cite{Guo2002}
\begin{equation}
	\label{eq54}
	F_i(\textbf{x},t)=\omega_i\left[\textbf{u}\cdot\nabla\rho+\frac{\textbf{c}_i\cdot\textbf{F}}{c_s^2}+\frac{\textbf{u}\nabla\rho:(\textbf{c}_i\textbf{c}_i-c_s^2\textbf{I})}{c_s^2}\right].
\end{equation}
$\textbf{S}^f=\text{diag}(s_0^f,s_1^f,s_2^f,s_3^f,s_4^f,s_5^f,s_6^f,s_7^f,s_8^f)$ is the diagonal relaxation matrix with $s_1^f$ and $s_2^f$ being related to the fluid viscosity as \cite{Chai2023}
\begin{equation}
	\label{eq55}
	\tau_f=0.5+\frac{\eta}{\rho c_s^2\Delta t}=\frac{1}{s_1^f}=\frac{1}{s_2^f},
\end{equation}
and the other parameters in $\textbf{S}^f$ are free. The velocity $\textbf{u}$ and hydrodynamic pressure $p$ are determined by the distribution function  $f_i(\textbf{x},t)$ as \cite{Guo2002}
\begin{equation}
	\label{eq56}
	\rho\textbf{u}=\sum_{i}\textbf{c}_if_i+\frac{1}{2}\Delta t\textbf{F},
\end{equation}
\begin{equation}
	\label{eq57}
	p=\frac{c_s^2}{1-\omega_0}\left[\sum_{i\neq0}f_i+\frac{\Delta t}{2}\textbf{u}\cdot\nabla\rho+\rho r_0(\textbf{u})\right].
\end{equation}

In summary, we propose an MRT-LB scheme for simulating two-phase EHD flows  induced by the Onsager-Wien effect, and the LB evolution equations in this scheme include Eq. (\ref{eq27}), Eq. (\ref{eq34}), Eq. (\ref{eq35}), Eq. (\ref{eq44}), and Eq. (\ref{eq51}). Before conducting numerical simulation with the present model, we would like to point out that since some derivatives appear in the current model, additional numerical schemes. In this work, the time derivative $	\partial_t(\phi\textbf{u})$ is evaluated by using the explicit Euler scheme \cite{SP2013,Liang PRE2018},
\begin{equation}
	\label{eq58}
	\partial_t(\phi\textbf{u})=\frac{\phi(t)\textbf{u}(t)-\phi(t-\Delta t)\textbf{u}(t-\Delta t)}{\Delta t}.
\end{equation}
The gradient operator is determined by employing a second-order isotropic central scheme \cite{SP2013,Liang PRE2018}
\begin{equation}
	\label{eq59}
	\nabla\chi(\textbf{x})=\sum_{i\neq0}\frac{\omega_i\textbf{c}_i\chi(\textbf{x}+\textbf{c}_i\Delta t)}{c_s^2\Delta t},
\end{equation} 
and the Laplace operator is computed by \cite{SP2013,Liang PRE2018}
\begin{equation}
	\nabla^2\chi(\textbf{x})=\sum_{i\neq0}\frac{2\omega_i\left[ \chi(\textbf{x}+\textbf{c}_i\Delta t)-\chi(\textbf{x})\right]}{c_s^2{\Delta t}^2},
\end{equation}
in which $\chi$ could be any physical variable. Another important aspect to consider is that since the current work focuses on the two-phase EHD flows, apart from the dimensionless parameters presented in Eq. (\ref{eq23}), the following parameters related to the ratio of different physical properties between two phases are also needed, 
\begin{equation}
	\label{eq60}
	\varepsilon_r=\frac{\varepsilon_l}{\varepsilon_v}, \quad  \mu_r=\frac{\mu_l}{\mu_v}, \quad \rho_r=\frac{\rho_l}{\rho_v},\quad \eta_r=\frac{\eta_l}{\eta_v},
\end{equation}
where the subscripts $v$ and $l$ denote the vapor phase and the liquid phase, respectively. Finally, to capture the physical variables within the phase interface, we employ a basic linear relationship for the order parameter, which is expressed as \cite{X He 1999}
\begin{equation}
	\label{eq61}
	\chi=\phi(\chi_l-\chi_v)+\chi_v.
\end{equation}
 
\section{Numerical validation}\label{section5}
In this section, we will conduct some numerical simulations to validate the developed LB scheme. First, we will examine the case of one-dimensional permeable electrodes submerged in a dielectric liquid. The numerical results indicate that the LB scheme is capable of simulating the EHD conduction problem induced by the Onsager-Wien effect. Next, to demonstrate that our numerical scheme can simulate two-phase EHD flows, we set the reaction term $R$ in Eq. (\ref{eq3}) and Eq. (\ref{eq4}) to zero. This simplifies the EHD flows to the traditional leaky model with surface charge convection \cite{suh2012}. As will be shown later, the resulting deformation factor aligns well with previous studies.

\subsection{One-dimensional permeable electrodes submerged in a dielectric liquid}\label{section5.1}

\begin{figure}[H]
	\centering
	\subfigure[]{\label{fig1a}\includegraphics[width=0.405\textwidth]{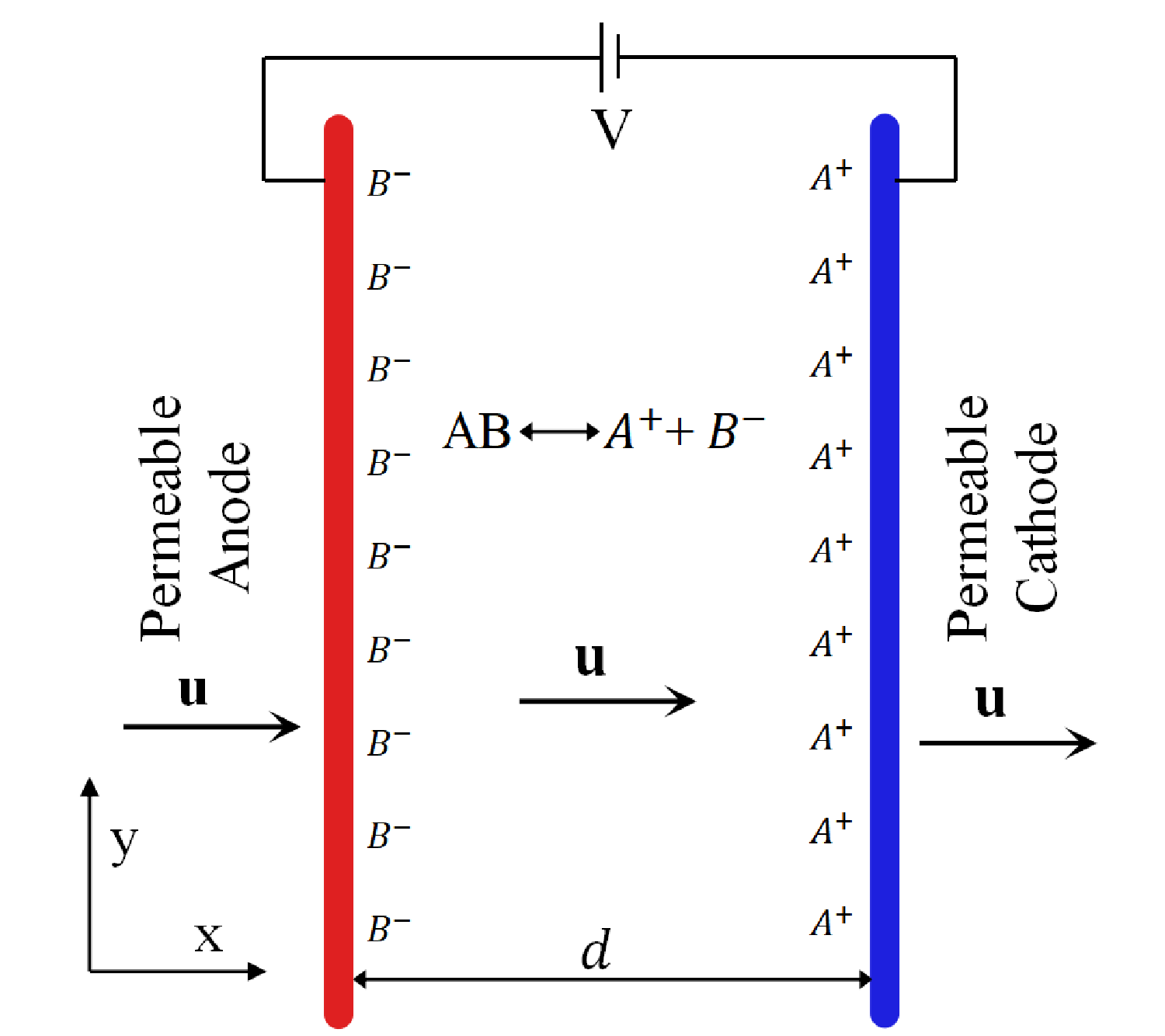}}
	\subfigure[]{\label{fig1b}\includegraphics[width=0.5\textwidth]{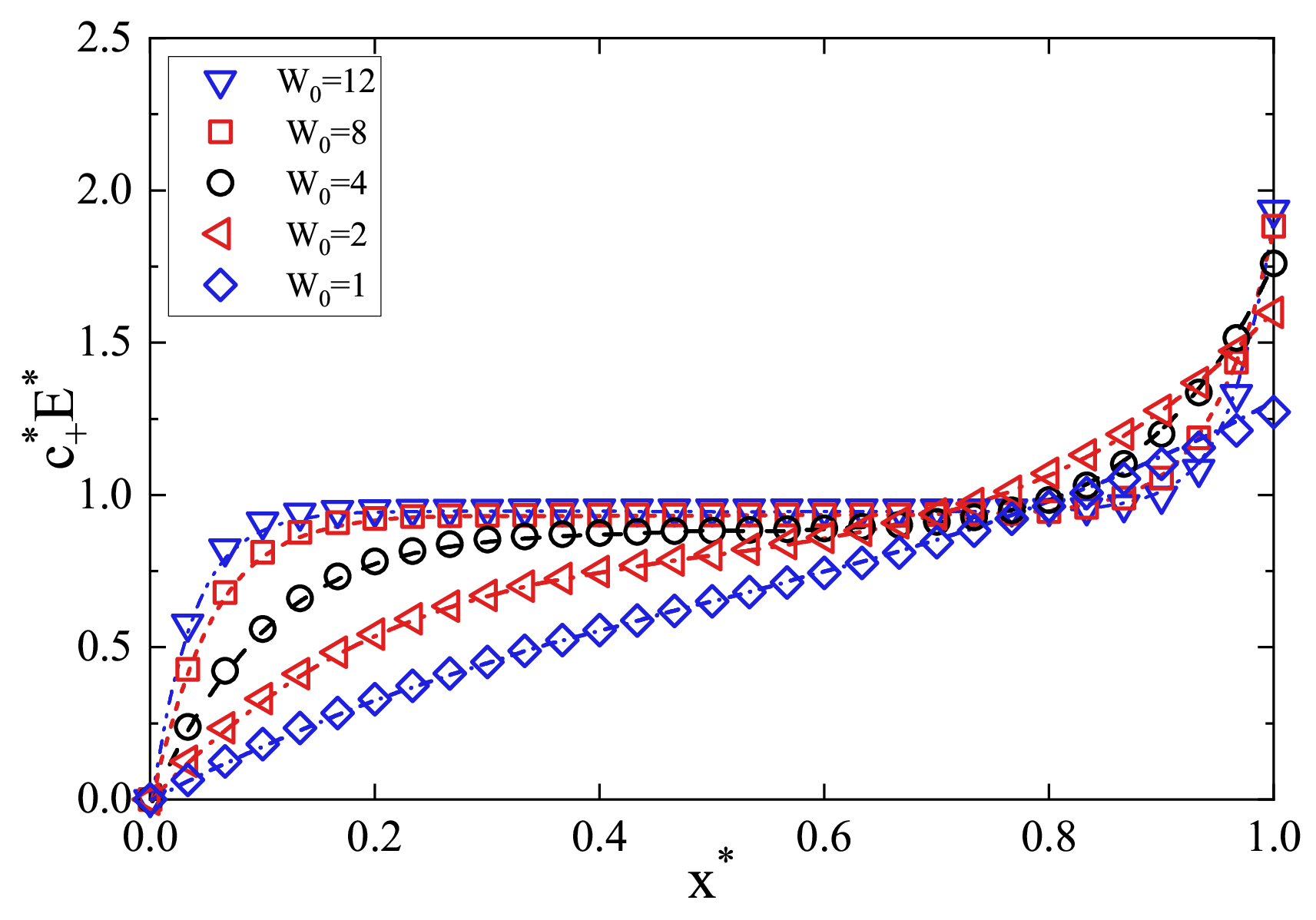}}
	\caption{(a) Schematic diagram of one-dimensional permeable electrodes submerged in a dielectric liquid, (b) the profiles of $c_+^*\textbf{E}^*$ along horizontal direction for different conduction numbers $W_0$, where the lines represent the analytical solutions in Ref \cite{Y.Feng2007}, and the symbols denote our numerical data.}
	\label{fig1}
\end{figure}

The configuration of one-dimensional permeable electrodes submerged in a dielectric liquid is depicted in Fig. \ref{fig1a}. The corresponding governing equations that describe this simplified problem could be expressed as \cite{Y.Feng2007}
\begin{equation}
	\begin{split}
		\frac{\partial(c_+\textbf{u}+c_+\textbf{E})}{\partial x}=&2W_0(1-c_+c_-),\,
		\frac{\partial(c_-\textbf{u}-c_-\textbf{E})}{\partial x}=2W_0(1-c_+c_-),\\
		\frac{\partial \textbf{E}}{\partial x}&=W_0(c_+-c_-),\quad\text{and}\quad
		\textbf{E} = -\frac{\partial \psi}{\partial x}.
	\end{split}
\end{equation}
In terms of the boundary conditions in the horizontal direction, they are defined as  
\begin{equation}
	\text{Anode : } \frac{\partial c_-}{\partial x}=0,\,\psi=1.0, \, c_+=0,\quad
	\text{Cathode : } \frac{\partial c_+}{\partial x}=0,\,\psi=0.0,\, c_-=0.
\end{equation}
For the vertical direction, the periodic boundary conditions are employed. Fig. \ref{fig1b} present the variations of  $c_+\textbf{E}$ for different conduction numbers $W_0$. It shows that the numerical results agree well with the analytical solutions, indicating that our numerical scheme is feasible for simulating the EHD conduction problem induced by the Onsager-Wien effect.

\subsection{Deformations of a leaky dielectric droplet}\label{section5.2}

\begin{figure}[H]
	\centering
	\subfigure[]{\label{fig2a}\includegraphics[width=0.49\textwidth]{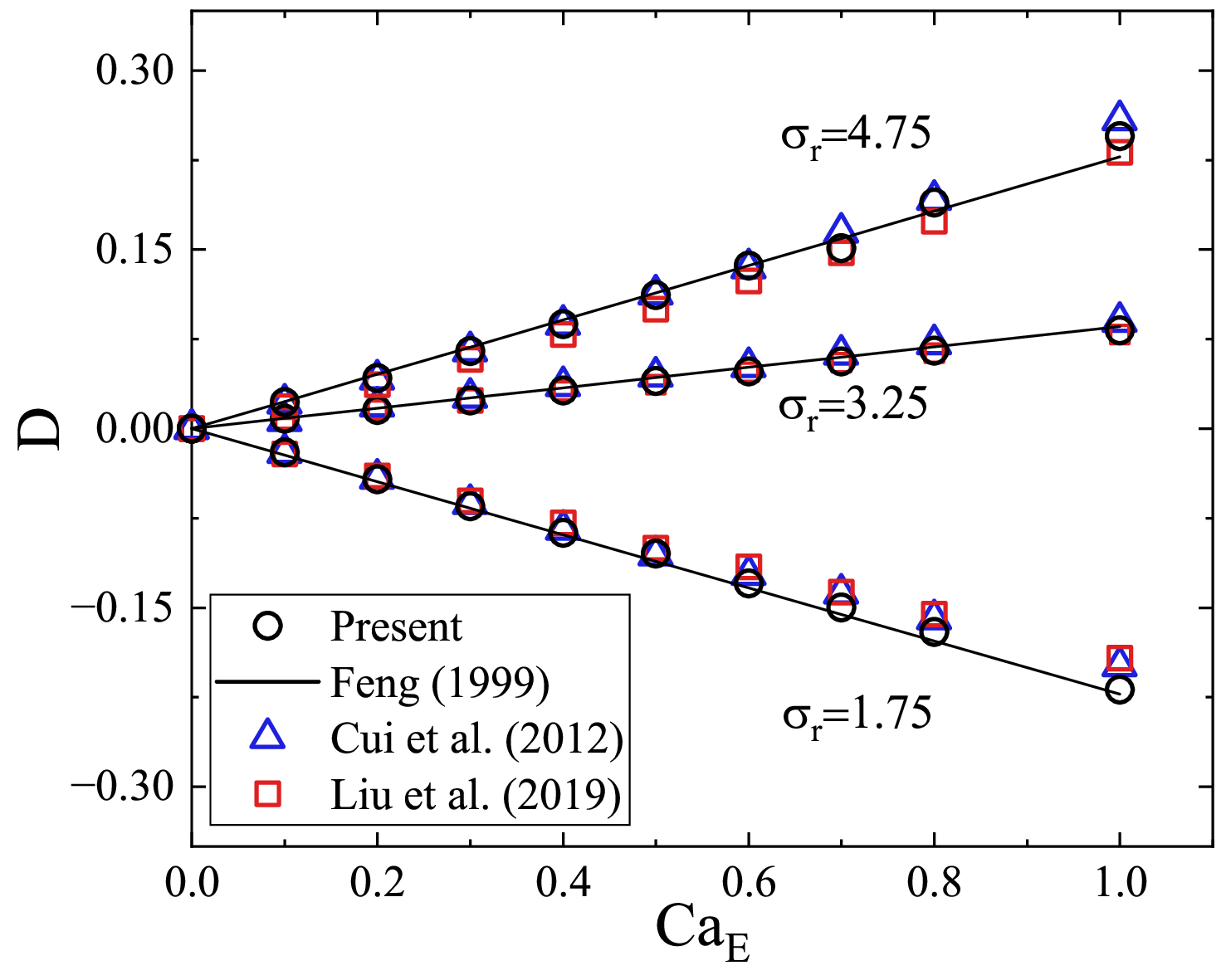}}
	\subfigure[]{\label{fig2b}\includegraphics[width=0.49\textwidth]{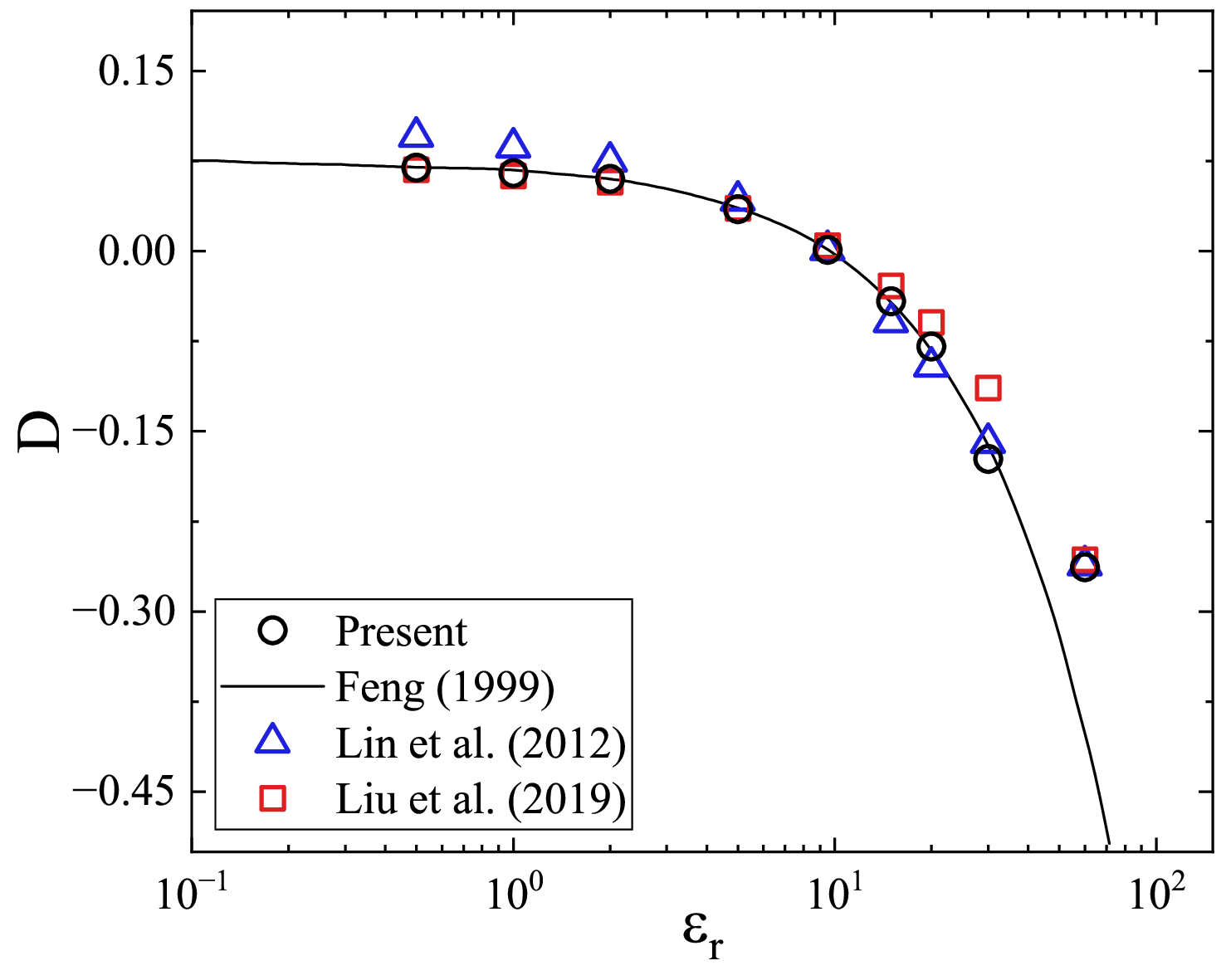}}
	\caption{Comparisons of the deformation factors $D$ for different electric capillary numbers $Ca_E$ at three typical  conductivity ratios $\sigma_r$ with $\varepsilon_r=3.5$ (a), and for different permittivity ratios $\varepsilon_r$ at $Ca_E=0.2$ and $\sigma_r=5.0$ (b).}
	\label{fig2}
\end{figure}

It is noted that when the reaction term $R$ in ionic transport equations is set to zero (i.e., $ R = 0.0 $), the governing equations correspond to the leaky dielectric model with surface charge convection \cite{suh2012}. In this scenario, choosing the deformations of a leaky dielectric droplet subjected to a uniform electric field to validate the developed numerical scheme is appropriate. The configuration of this classic two-phase EHD problem can be easily found elsewhere \cite{Feng1999,Liu2019,Cui2019}, and the state of the droplet is characterized by the deformation factor
\begin{equation}
	\label{eq63}
	D=\frac{L-H}{L+H},
\end{equation} 
in which $L$ denotes the length of the droplet along the axis parallel to the electric field, while $H$ represents the length of the droplet along the axis perpendicular to the electric field direction. Fig. \ref{fig2} shows the comparisons of the $D$ profile between the numerical result and previous data for different electric capillary numbers $Ca_{E}$, conductivity ratios $\sigma_r=\sigma_l/\sigma_v$, and permittivity ratios $\varepsilon_r$. It can be seen that our numerical results agree well with previous works, which suggests that the present model can achieve satisfying accuracy in solving two-phase EHD problems.      

\section{Results and discussion}\label{section6}
\begin{figure}[H]
	\centering
	\includegraphics[width=0.5\textwidth]{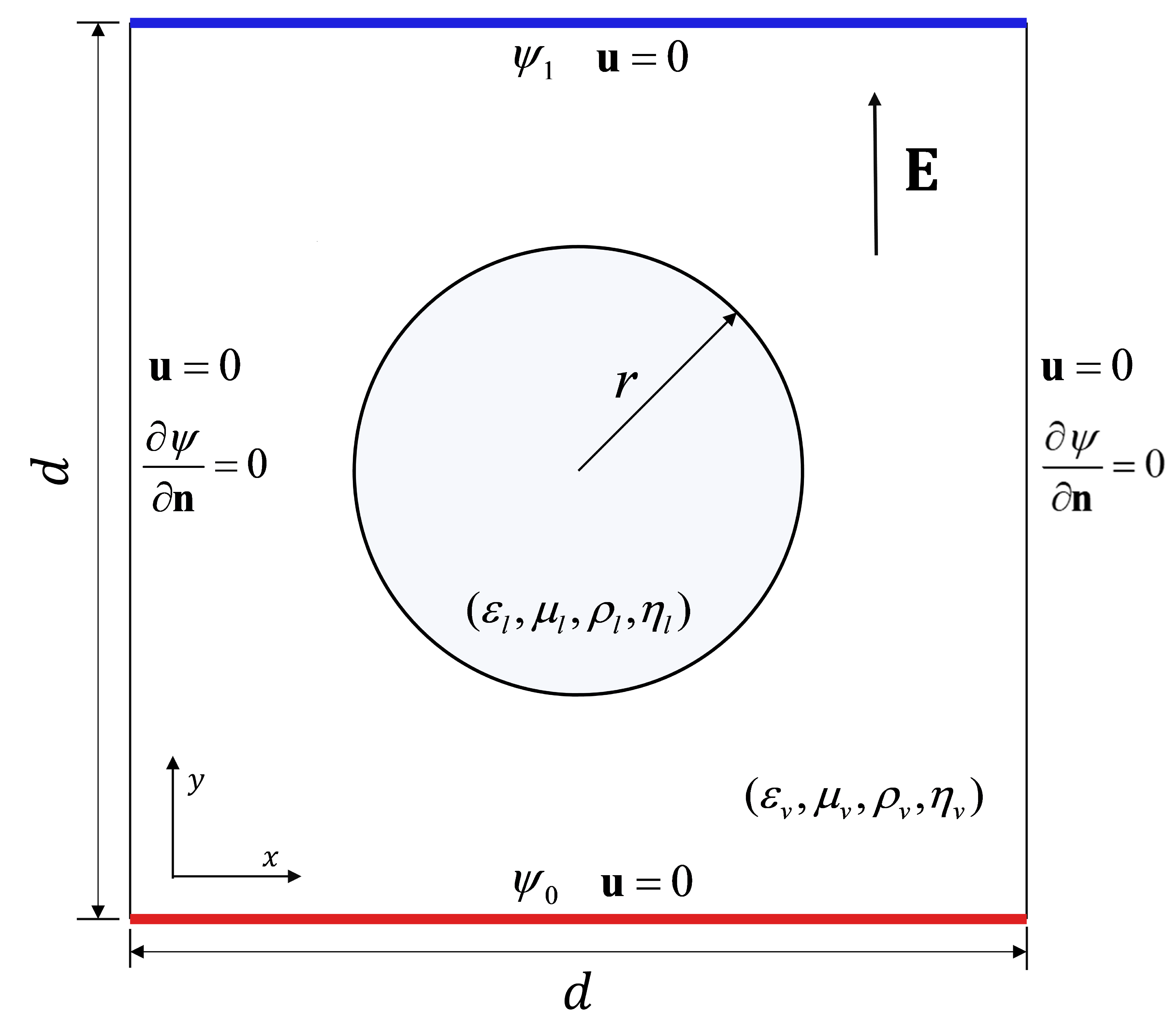}
	\caption{Schematic representation of a droplet suspended in a dielectric fluid, subjected to a uniform electric field. The subscripts $v$ and $l$ denote the external fluid and internal fluid, respectively.}
	\label{fig3}
\end{figure}

In this section, the numerical results for droplet deformation under the EHD conduction phenomenon induced by the Onsager-Wien effect will be discussed, with the configuration depicted in Fig. \ref{fig3}. At the beginning, we consider a droplet with an initial radius $r$ suspended in a dielectric fluid, positioned within a square enclosure that contains two parallel planar electrodes. Subsequently, a uniform electric field $\textbf{E}$ is exerted in the vertical direction, creating an electric potential difference $\Delta\psi=\psi_0-\psi_1$ between the two electrodes. In such a case, the droplet undergoes a deformation under the electric field. In addition, based on the previous work \cite{SIJeong 2003,WangAMM2021}, the boundary conditions can be set as
\begin{subequations}
	\label{eq64}
	\begin{align}
		\text{Anode : }\, c_+=0,\,&\textbf{u}=0,\,\psi=\psi_0,\,\text{and}\,\frac{\partial c_-}{\partial\textbf{n}}=0, \quad
		\text{Cathode : }\, c_-=0,\,\textbf{u}=0,\, \psi=\psi_1,\, \text{and}\,\frac{\partial c_+}{\partial\textbf{n}}=0,\\
		&\text{The other boundaries : }\, \frac{\partial c_-}{\partial\textbf{n}}=0,\, \frac{\partial c_+}{\partial\textbf{n}}=0,\,\frac{\partial\psi}{\partial\textbf{n}}=0\, \text{and}\,\textbf{u}=0,
	\end{align}
\end{subequations}
in which $\textbf{n}$ is the normal vector of the wall. Further, the main parameters used in the current work are listed in Tab. \ref{table1}, and unless otherwise specified, the following fixed values are employed in the simulations: $d = 2.0 \times 10^{-3}$ (m), $r=d/4$, $\Delta\psi = 1.0$ (KV), $\rho_r=2.0$, $\varepsilon_r = 0.5$, $\mu_r = 5.0$, $m = 1.0$, $Ca_E = 0.241$, $\zeta = 0.001$, $M_{\phi}=0.1$, $Cn = 0.015$. Moreover, the other parameters in the diagonal relaxation matrix are set to 1.0  \cite{Chai2023}. Finally, we will perform a detailed discussion on the effects of geometric scale ($d$), applied voltage ($\Delta\psi$), permittivity ratio ($\varepsilon_r$), and ionic mobility ratio ($\mu_r$) on the droplet deformation and charge distribution.

\begin{figure}[H]
	\centering
	\subfigure[]{\label{fig10a}\includegraphics[width=0.314\textwidth]{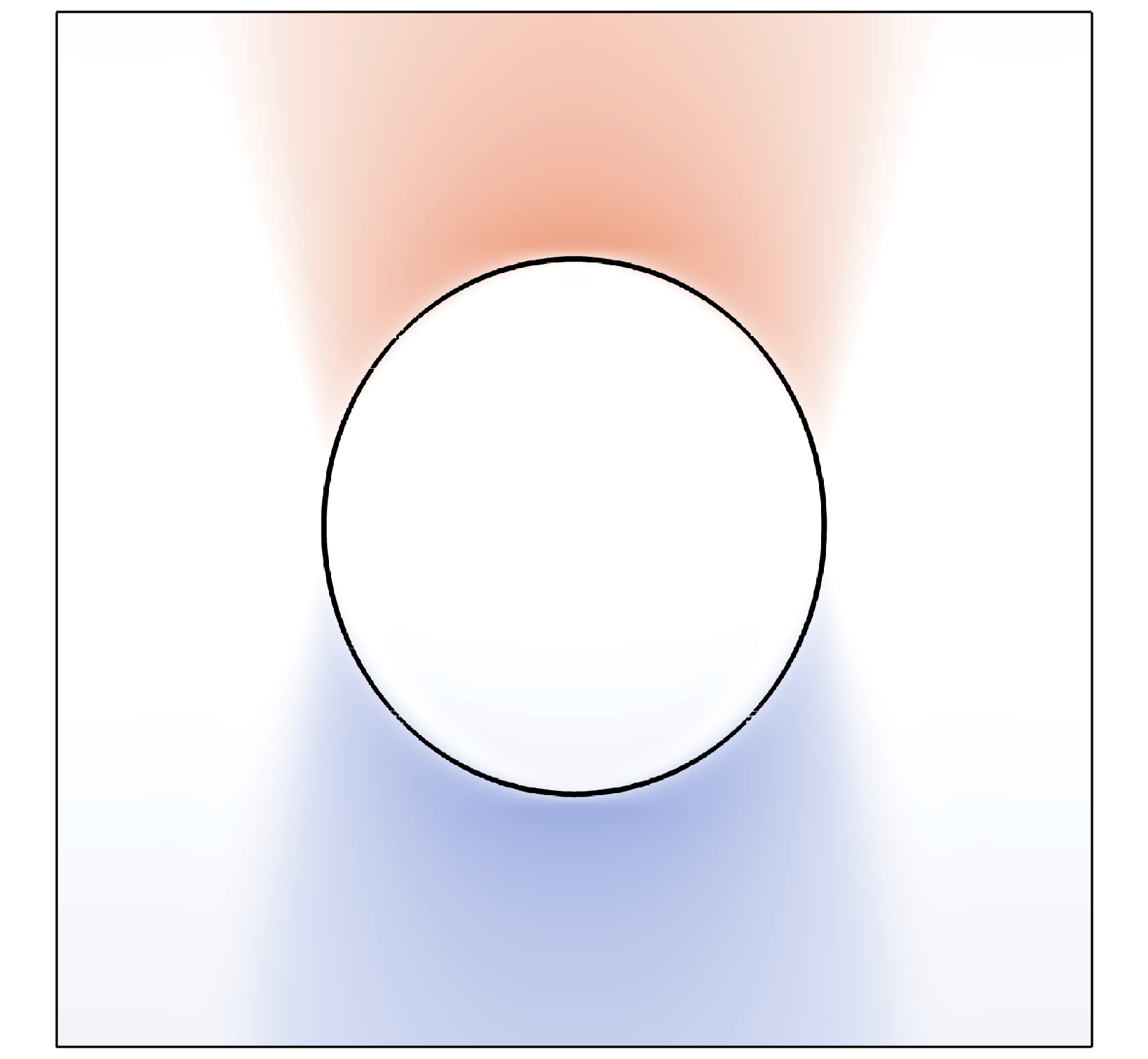}}
	\subfigure[]{\label{fig10b}\includegraphics[width=0.314\textwidth]{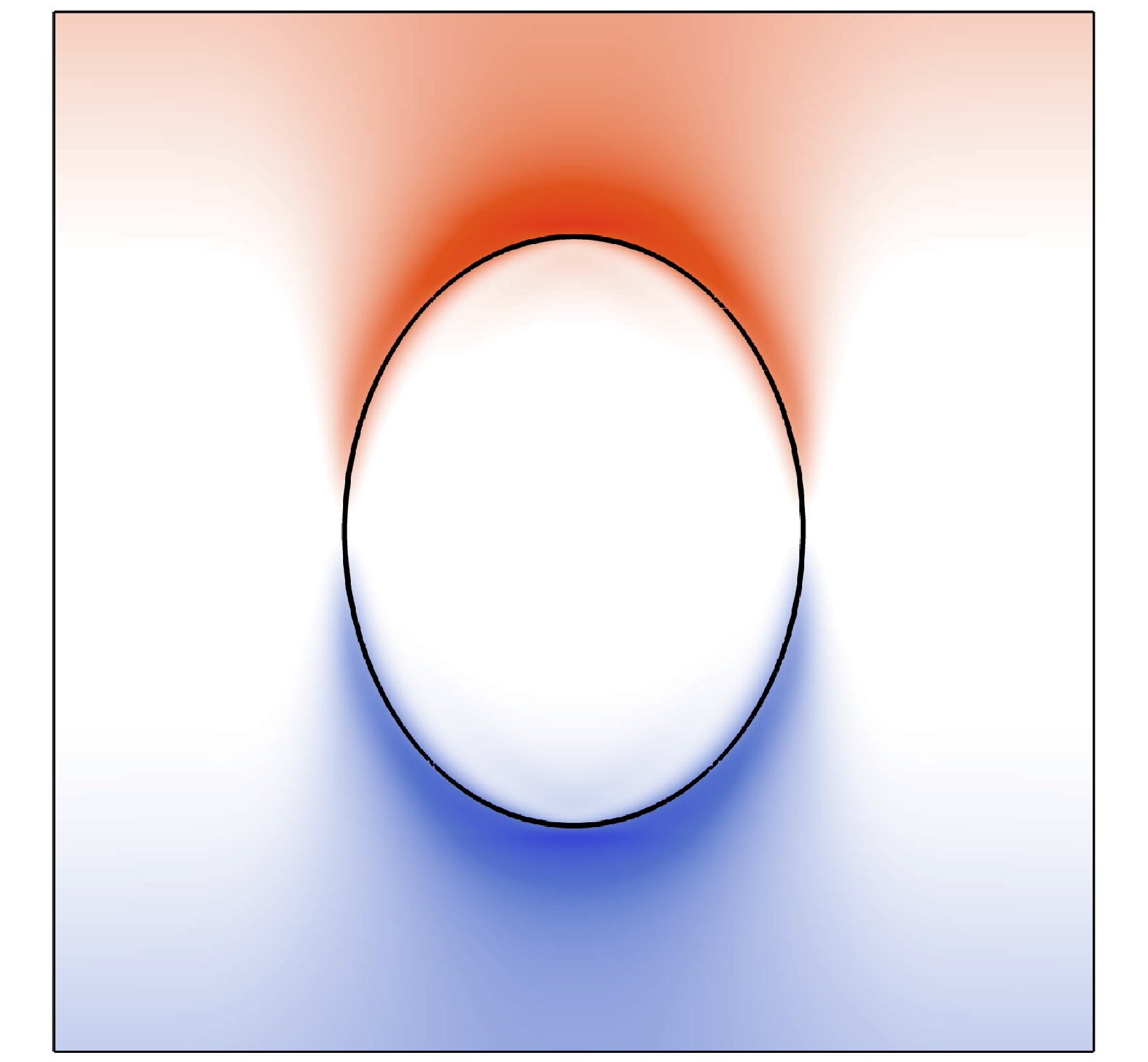}}	
	\subfigure[]{\label{fig10c}\includegraphics[width=0.351\textwidth]{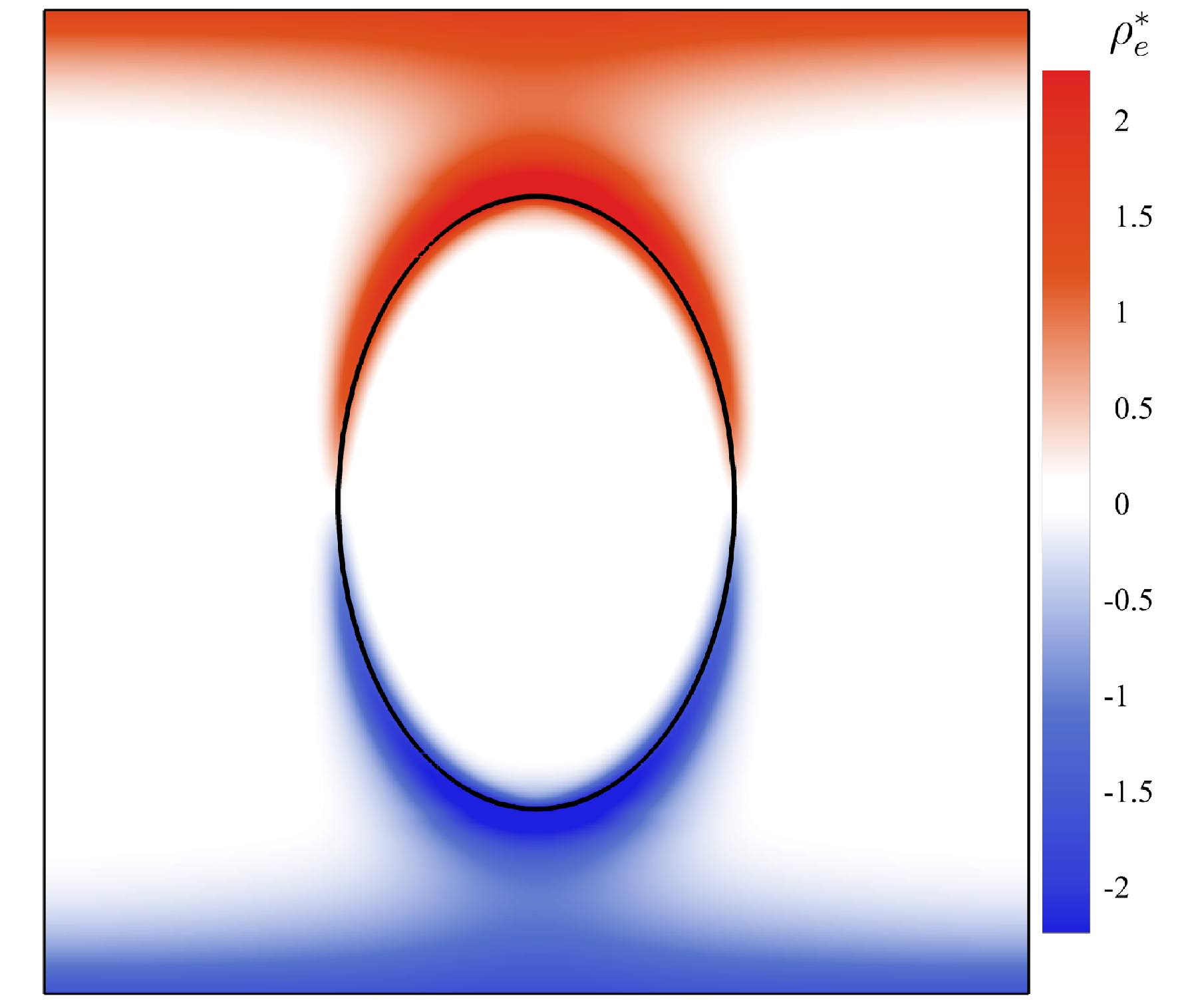}}	\\
	\caption{The charge distribution (colored regions) and the droplet profile (black lines) for three different reference lengths: (a) $d=1.0$ mm, $W_0=1.42233$; (b) $d=2.0$ mm, $W_0=5.6893$; (c) $d=4.0$ mm, $W_0=22.7572$.}
	\label{fig10}
\end{figure}

We first investigate the influence of the reference length $d$ on charge distribution and droplet deformation. Fig. \ref{fig10} presents the charge distribution and droplet shape at three different reference lengths (i.e., $d=1.0$ mm, $d=2.0$ mm, and $d=3.0$ mm). As shown in the figure, for each reference length, positive charges accumulate at the upper part of the droplet, while negative charges concentrate at the bottom. In addition, heterocharge layers are always present near the electrodes, which are not absent in the leaky dielectric model \cite{Liu2019,Cui2019,Luo POF 2020}. Furthermore, as the reference length $d$ increases, the thickness of the heterocharge layer around the electrodes gradually decreases. Meanwhile, the concentration of diffuse space charge (also known as charge cloud) \cite{D.A.Saville1997,Bay1990} between the droplet interface and the electrode diminishes, and the droplet shape transitions from round to slender.

To explain these phenomena, it is necessary to determine the operating state of the current EHD conduction pump, which depends on the conduction number $W_0$ \cite{Atten2003,VazquezPOF2019}. The conduction number $W_0$ is defined as $W_0=\tau_K/2\tau_{\sigma}^0$, where $\tau_K=d^2/\overline{\mu}\Delta\psi$ represents the ionic transit time, which is the typical duration that ions take to move from one electrode to another. $\tau_\sigma^0=\varepsilon/\sigma_0$ denotes the ohmic time, which represents the typical duration for ions to recombine within the bulk \cite{VazquezPOF2019, WangAMM2021}. Typically, ions travel a certain distance before recombining, which is expressed as $\iota_H \sim d/(2W_0)$. In addition, when $W_0 \gg 1.0$, we observe that $\tau_K \gg \tau_\sigma^0$ and $\iota_H\ll d$, meaning that ions have sufficient time to recombine within the bulk and form an electroneutral region. In this case, the system is operating in the ohmic state. While when $W_0 \ll 1.0$, we have $\tau_K \ll \tau_\sigma^0$ and $\iota_H\gg d$, meaning that ions do not have enough time to recombine, which prevents the formation of an electroneutral region, and the system operates in the saturation state \cite{VazquezPOF2019, WangAMM2021}.

\begin{figure}[H]
	\centering
	\subfigure[]{\label{fig11a}\includegraphics[width=0.324\textwidth]{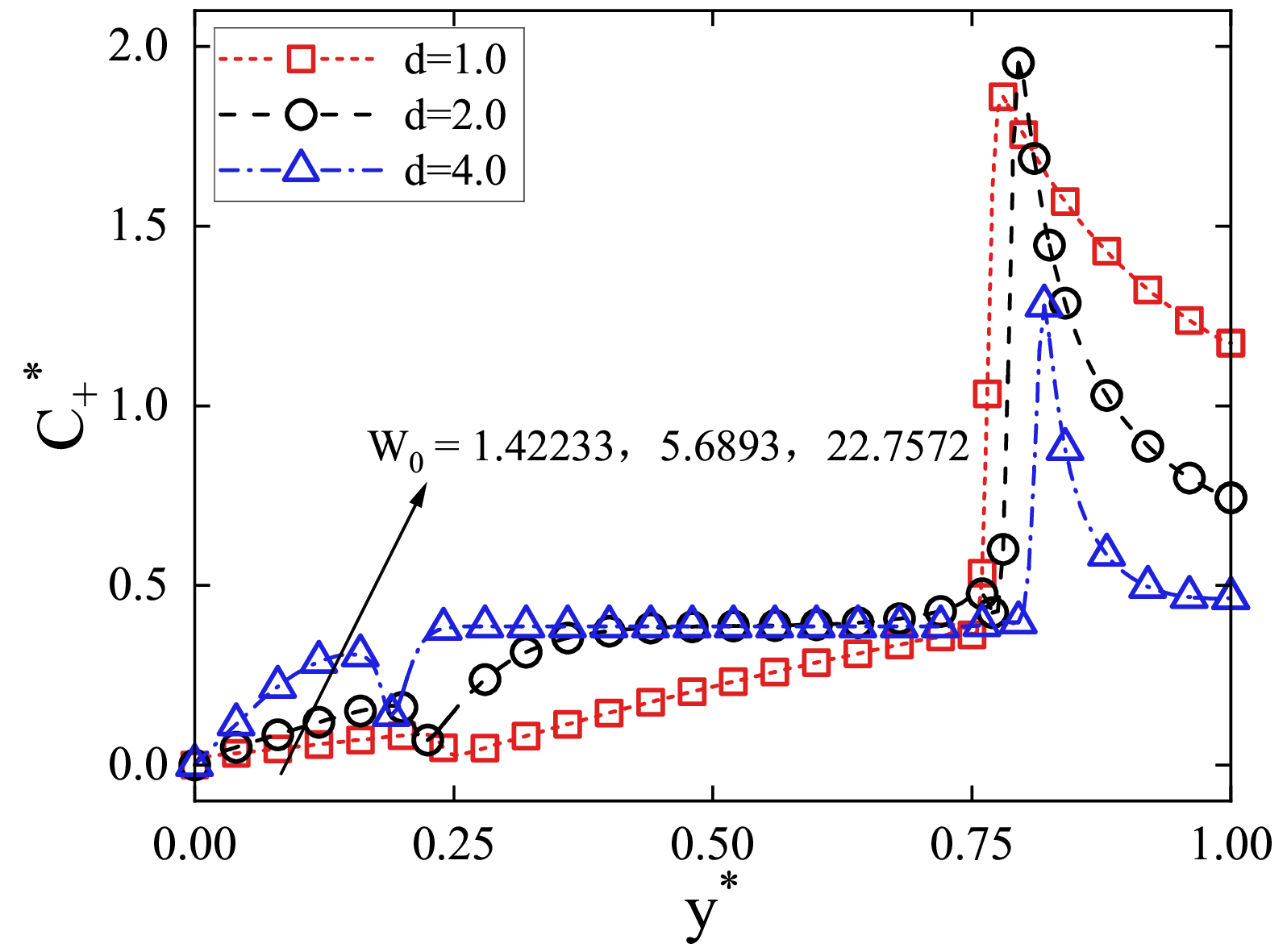}}	
	\subfigure[]{\label{fig11b}\includegraphics[width=0.323\textwidth]{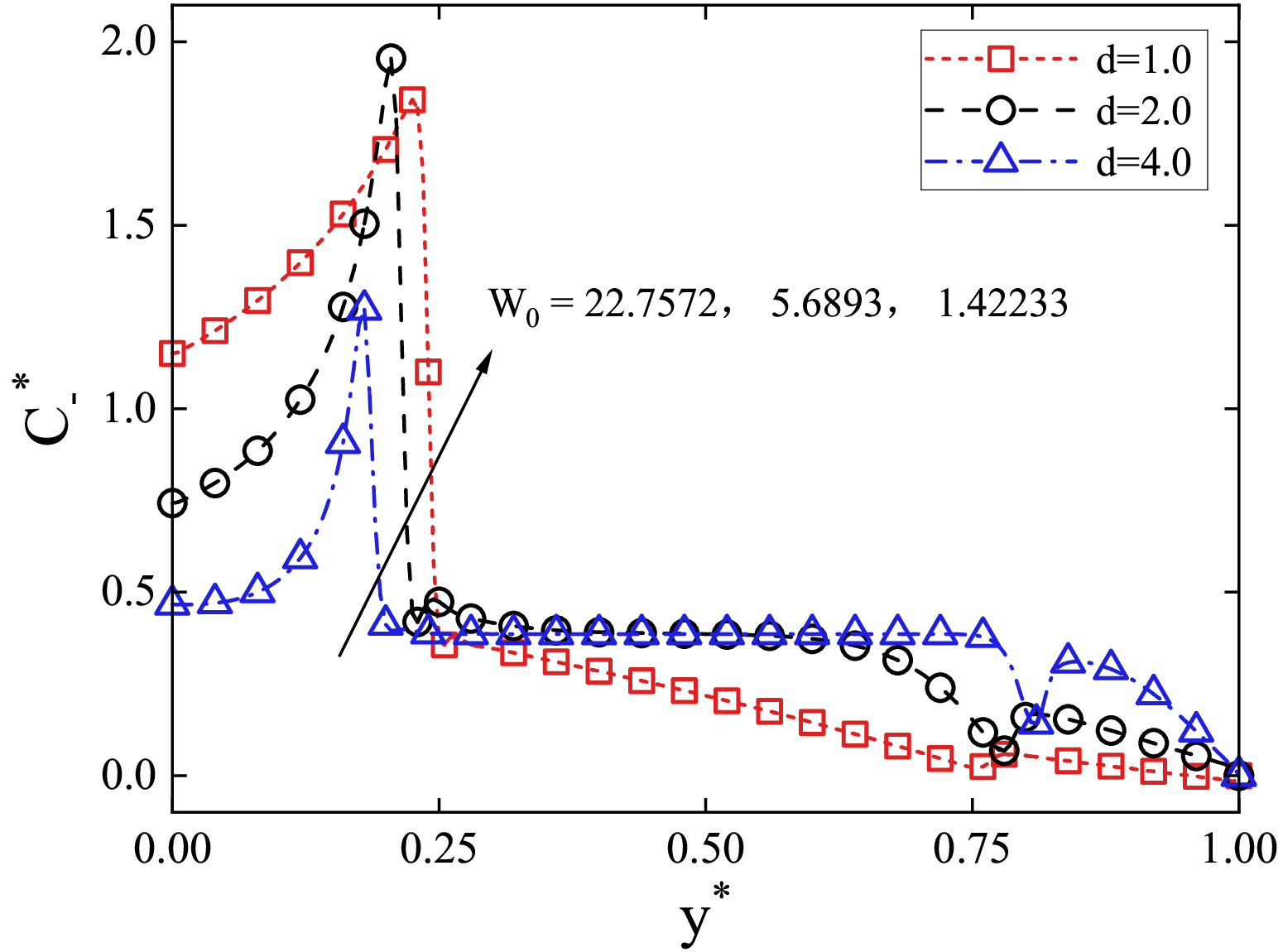}}	
	\subfigure[]{\label{fig11c}\includegraphics[width=0.326\textwidth]{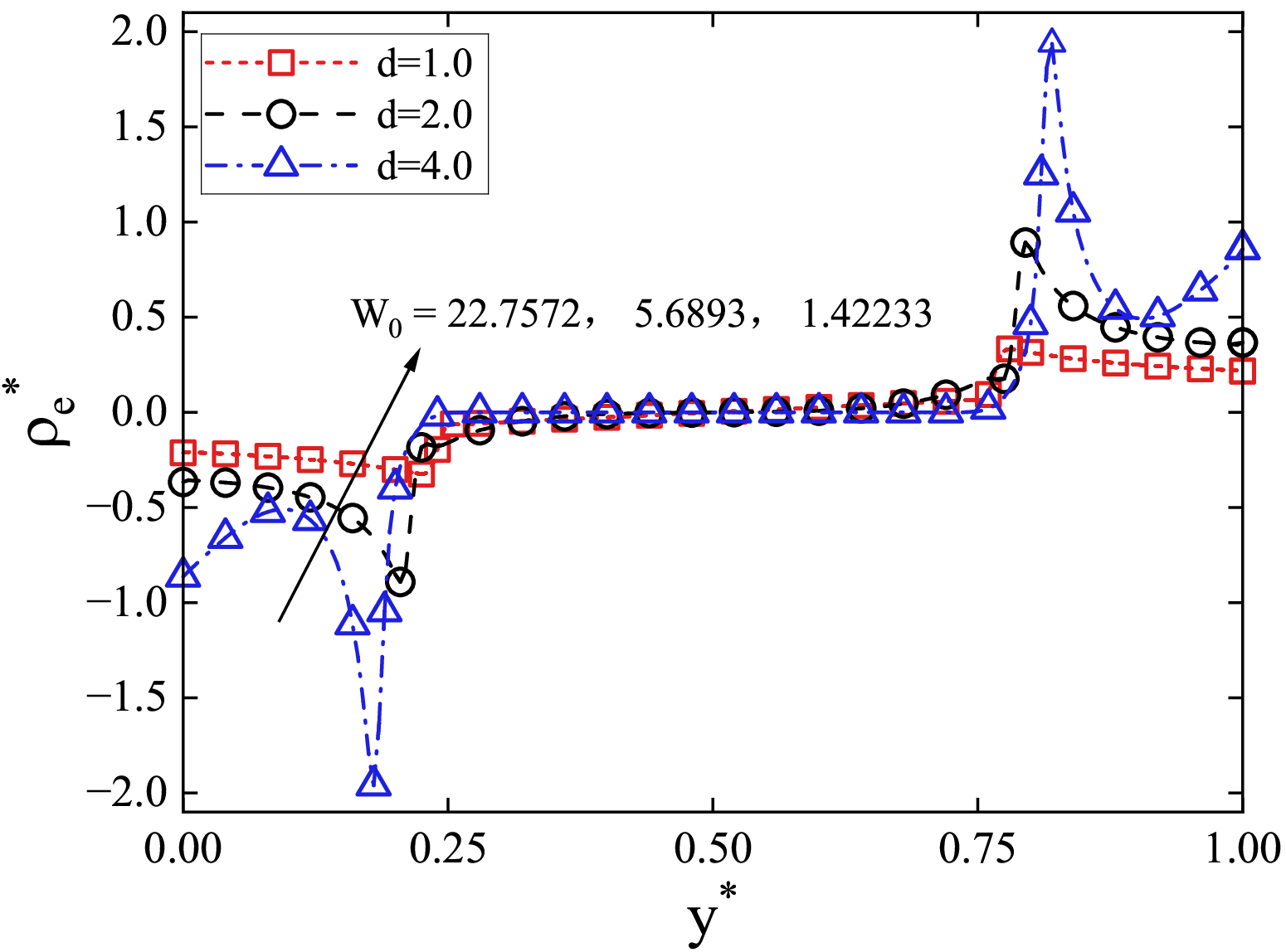}}	\\
	\caption{The distributions of cations (a), anions (b), and charge (c) at $x=0.5d$ within the cavity for three different reference lengths.}
	\label{fig11}
\end{figure}

\begin{figure}[H]
	\centering
	\subfigure[]{\label{fig12a}\includegraphics[width=0.3125\textwidth]{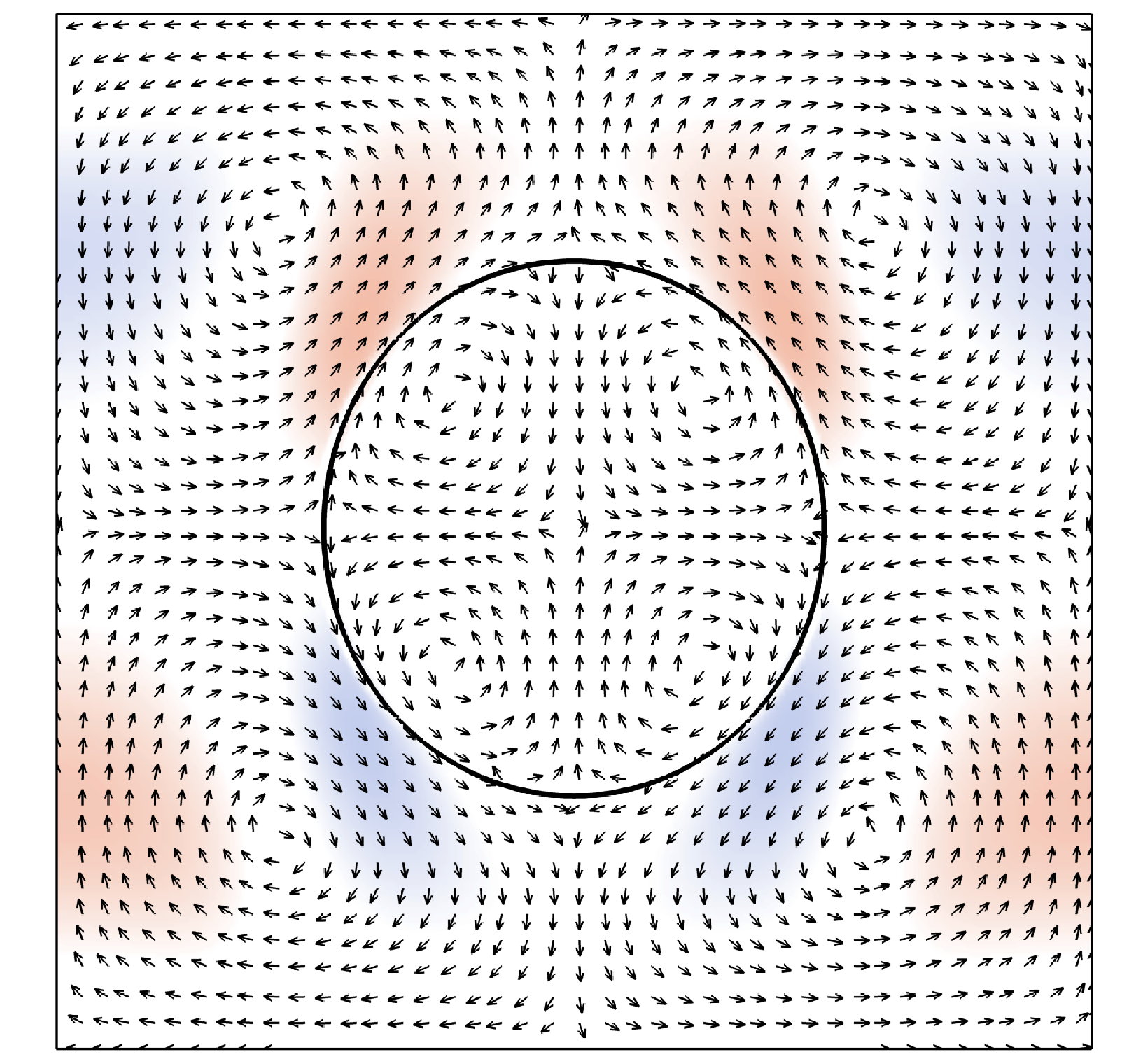}}
	\subfigure[]{\label{fig12b}\includegraphics[width=0.312\textwidth]{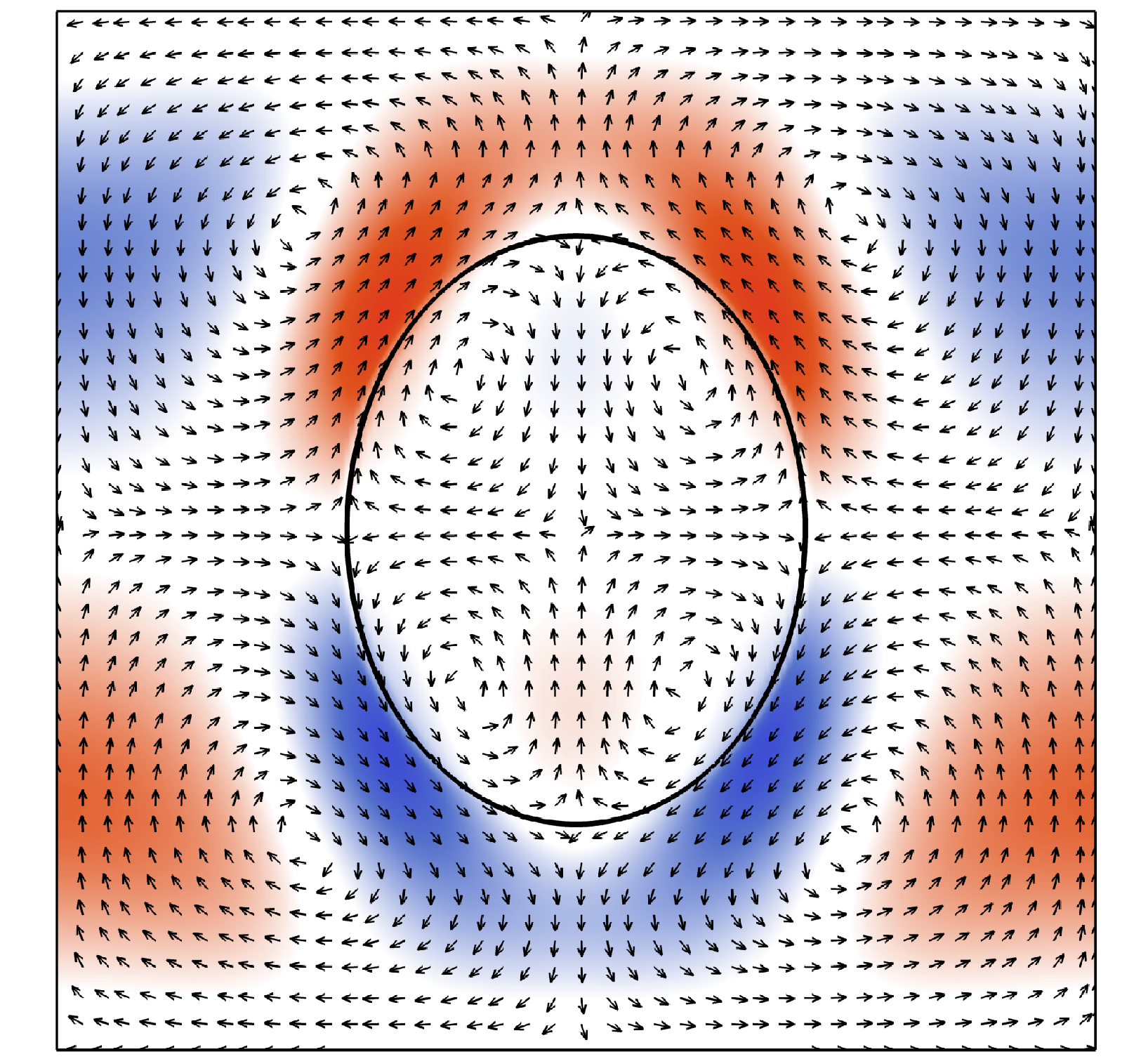}}
	\subfigure[]{\label{fig12c}\includegraphics[width=0.36\textwidth]{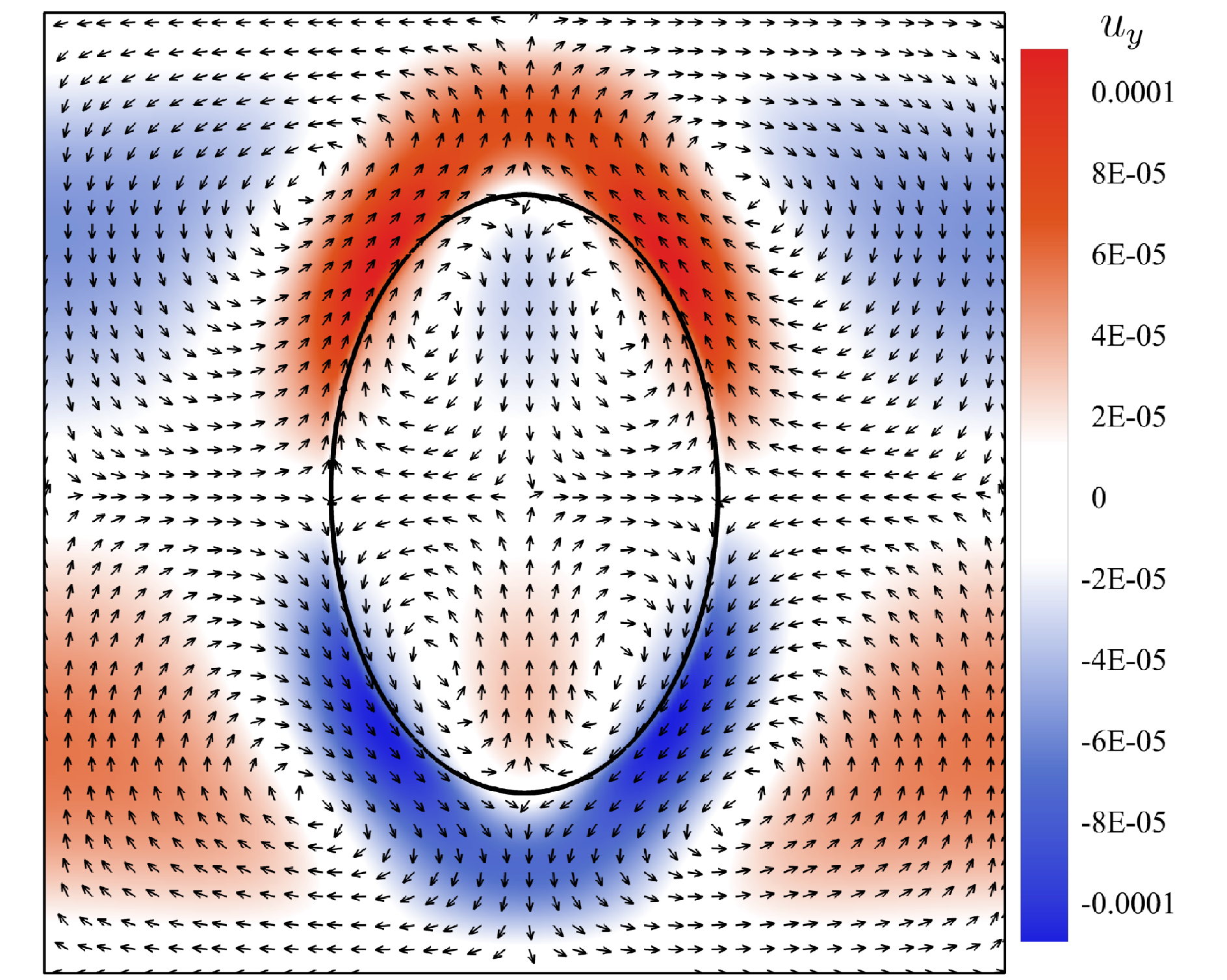}}	\\
	\caption{The distributions of the velocity vectors (black arrows), vertical velocity $u_y$ (colored parts), and the profile of the droplet (black lines) for three different reference lengths: (a) $d=1.0$ mm; (b) $d=2.0$ mm; (c) $d=4.0$ mm.}
	\label{fig12}
\end{figure}

Returning to Fig. \ref{fig10}, one can observe that the conduction numbers $W_0$ corresponding to the three different lengths $d$ are all greater than 1.0, indicating that the system works in the ohmic state. In this state, the ionic transit time $\tau_K$ is longer than the ohmic time $\tau_\sigma^0$, implying that the ions within the droplet have sufficient time to recombine, leading to the formation of an electroneutral region. To illustrate this further, Fig. \ref{fig11} presents the distribution of positive and negative ions, as well as the charge distribution, along the vertical direction at the center of the cavity for the three different reference lengths. It is noted that the permittivity ratio $\varepsilon_r=0.5$, since the fluid with a higher permittivity exhibits a stronger polarization response to the electric field, which results in a stronger electric field within the droplet. At the same time, since the ionic mobility ratio $\mu_r=5.0$, leads to higher ionic mobility inside the droplet. Under these conditions, the cations are attracted to the cathode, leading to their predominant accumulation at the upper droplet interface and cathode (see Fig. \ref{fig11a}). Conversely, anions are drawn toward the anode, resulting in their primary distribution at the lower droplet interface and anode (see Fig. \ref{fig11b}). This ion distribution further results in the charge distribution illustrated in Fig. \ref{fig11c}. It is also observed that as $d=2$ mm and $d=4$ mm, an electroneutral region forms inside the droplet. However, when $d=1.0$ mm, although the system is in the ohmic state, the relatively small conduction number $W_0$ means the ionic transit time $\tau_K$ is not significantly different from the ohmic time $\tau_\sigma^0$. Additionally, the droplet interface affects ionic transit and prevents ions from recombining within the droplet. Consequently, an electroneutral region does not form within the droplet, and the system exhibits behavior characteristic of the saturation state. Further observation of Fig. \ref{fig10} shows that as $d$ increases, both the thickness of the heterocharge layer and the concentration of the charge cloud decrease. In fact, the thickness of the heterocharge layer is defined as $\iota = \iota_H/d$, where $\iota_H$ represents the characteristic thickness. It is evident that $\iota$ exhibits an inverse relationship with the conduction number $W_0$. Therefore, as the reference length increases from $d=1.0$ mm to $d=4.0$ mm, the conduction number $W_0$ increases, leading to a reduction in the thickness of the heterocharge layer $\iota$. With the increase in $W_0$, more charge accumulates at the droplet interface and near the electrodes (see Fig. \ref{fig11}), which causes the concentration of the charge cloud between the droplet interface and the electrode to decrease. The numerical results for different reference lengths $d$ indicate that, regardless of whether an electroneutral region exists inside the droplet, charge always accumulates near the electrodes and the droplet interface. In this case, an electric field force arises near the electrodes and the droplet interface, which further interacts with the charges and induces fluid flow both inside and outside the droplet. To better understand this phenomenon, Fig. \ref{fig12} presents velocity vectors and vertical velocity $u_y$ distribution for different reference lengths. It can be seen that, for all values of $d$, there are always four vortices, with opposite and symmetric flow directions inside and outside the droplet. Moreover, as the reference length $d$ increases, the charge accumulated at the droplet interface also increases (see Fig. \ref{fig10} or Fig. \ref{fig11c}), which intensifies the electric field force exerted on the droplet interface. As a result, the droplet undergoes greater deformation and generates higher fluid velocities (see Fig. \ref{fig12}).

After discussing the impact of the reference length $d$, we now investigate the influence of the applied voltage $\Delta\psi$ on charge distribution and droplet deformation. To analyze this effect more comprehensively, we introduce a new dimensionless parameter $\beta$, which depends solely on the reference length $d$. Specifically, when $\beta$ is small (i.e., $d$ is small), changes in the applied voltage $\Delta\psi$ may cause the system to transition between the ohmic and saturation states. While, when $\beta\textgreater 1.0$, changes in the applied voltage $\Delta\psi$ do not affect the system state, which remains in the ohmic state \cite{VazquezPOF2019,WangAMM2021}. In this context, two reference lengths are fixed at $d=2$ mm and $d=8$ mm, corresponding to $\beta=0.5643$ and $\beta=1.1286$, respectively, and the system is analyzed by varying $\Delta\psi$ from 0.5 KV to 10.0 KV.

\begin{figure}[H]
	\centering
	\subfigure[]{\label{fig14a}\includegraphics[width=0.485\textwidth]{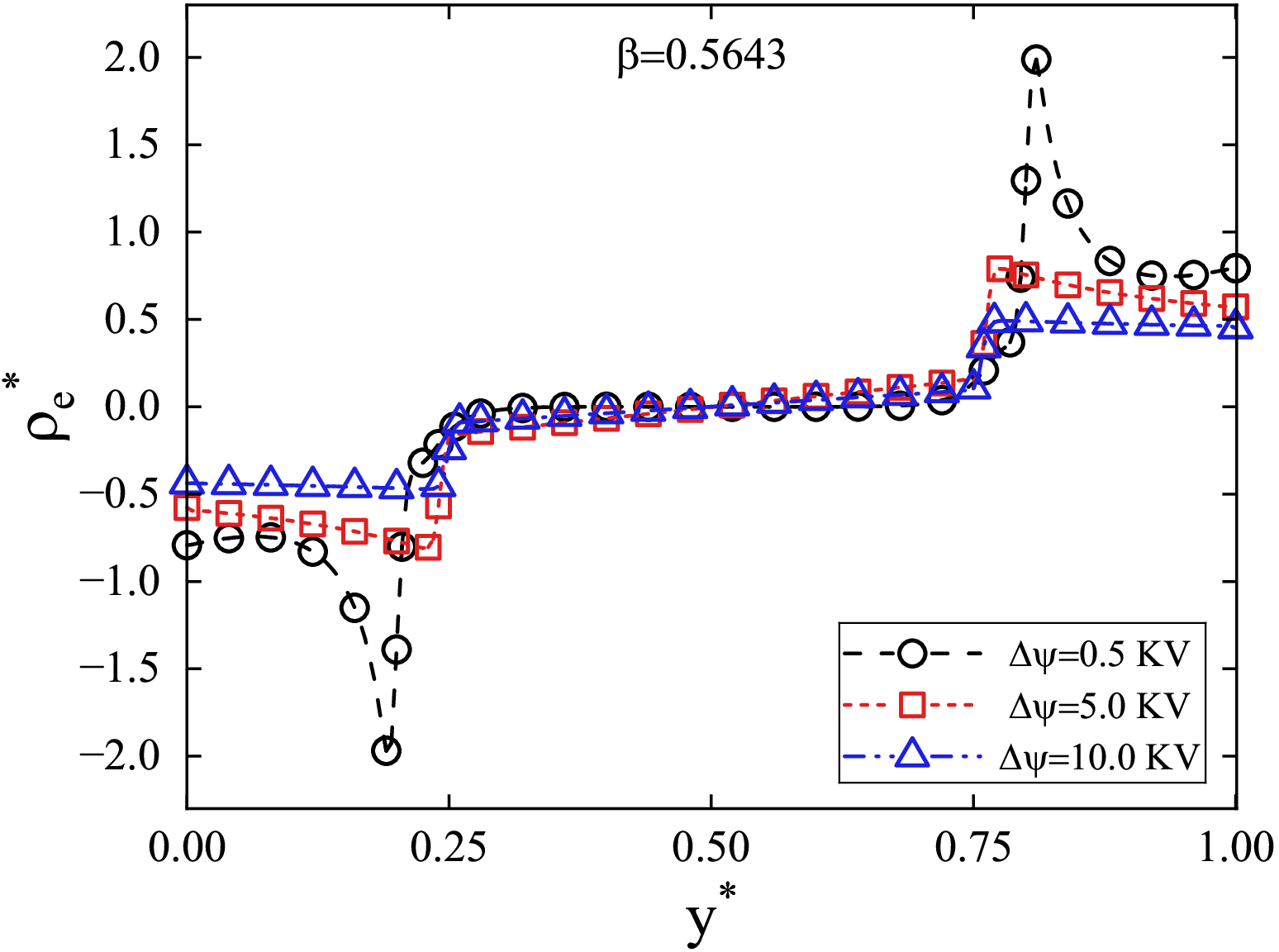}}
	\subfigure[]{\label{fig14b}\includegraphics[width=0.495\textwidth]{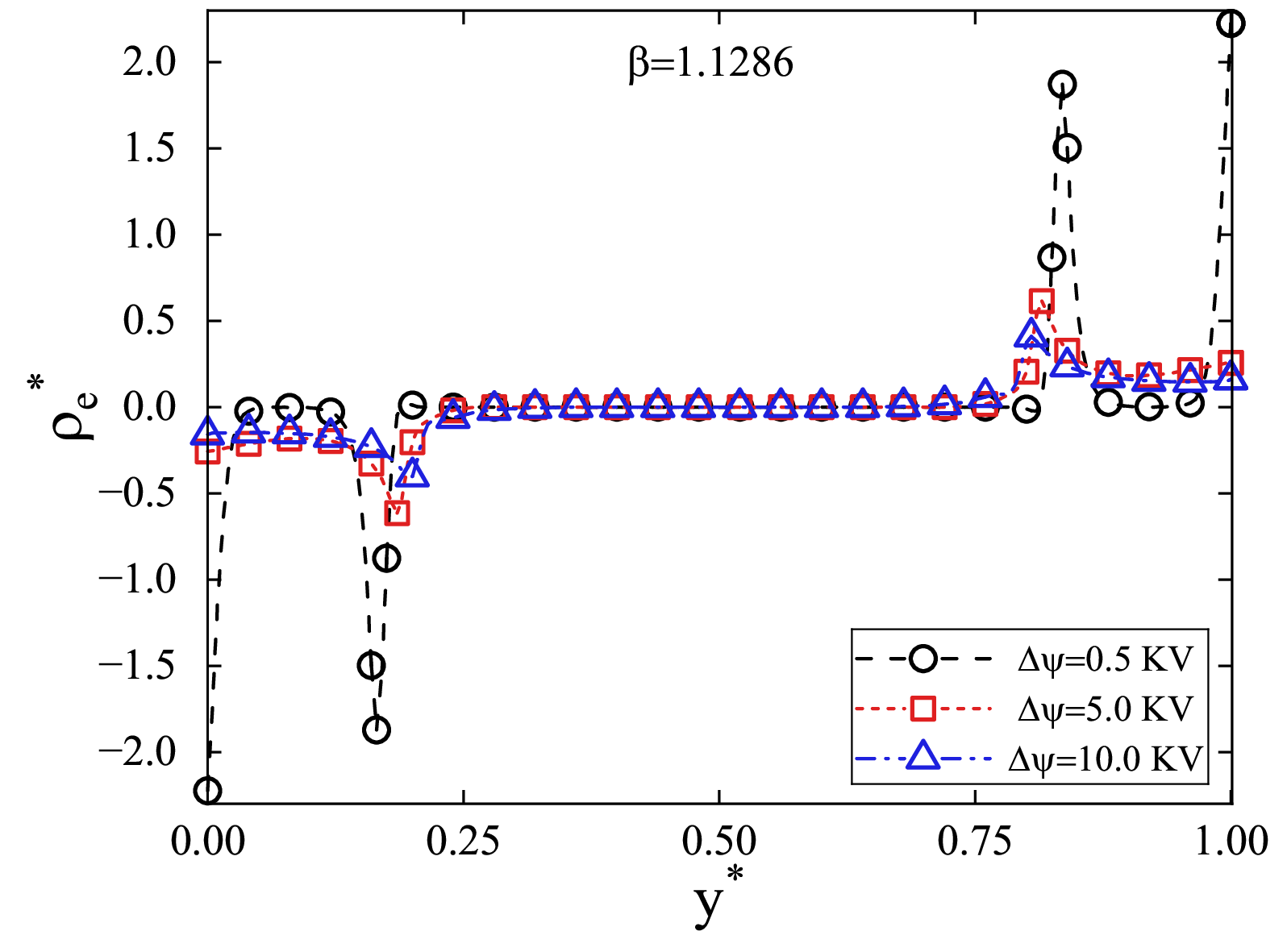}}
	\caption{The distributions of the charge in the cavity at $x=0.5d$ for $d=2.0$ mm (a) and $d=8.0$ mm (b).}
	\label{fig14}
\end{figure}

To visualize the numerical results, Fig. \ref{fig14} shows the charge distribution curves along the vertical center of the square cavity for two reference lengths, in which the parameter $\beta$ is also included. It has been observed that for $\beta=0.5643$, as the applied voltage $\Delta\psi$ from 0.5 KV to 10.0 KV, the electroneutral region inside the droplet gradually disappears, which signifies that the system transitions from the ohmic state to the saturation state. In fact, under this condition, the reference length $d=2$ mm is relatively small, and the conduction number $W_0$ decreases as the applied voltage $\Delta\psi$ increases. Consequently, the ion transit time $\tau_K$ approaches or falls below the ohmic time $\tau_\sigma^0$, which prevents ion recombination and leads to the gradual disappearance of the electroneutral region inside the droplet. Meanwhile, as the conduction number $W_0$ decreases, the charge accumulation at the droplet interface relatively decreases, while the thickness of the heterocharge layer $\iota$ relatively increases. This results in a higher concentration of charge cloud between the droplet interface and the electrodes. Therefore, for smaller $\beta$, an increase in $\Delta\psi$ gradually drives the system toward the saturation state. However, when $\beta=1.1286$, an electroneutral region remains inside the droplet even at $\Delta\psi= 10.0$ kV (see Fig. \ref{fig14b}). At this point, the reference length $d=8$ mm is relatively large, and despite the applied voltage $\Delta\psi$ increases to 10.0 kV, the ion transit time $\tau_K$ remains longer than the ohmic time $\tau_\sigma^0$. This allows ions to have sufficient time to fully recombine within the droplet, thereby maintaining the system in the ohmic state. Additionally, it is noteworthy that when the applied voltage $\Delta\psi=0.5$ kV, the charge cloud concentration between the heterocharge layer and the droplet interface reduces to zero. At this stage, the conduction number $W_0$ is large, and the ion transit time $\tau_K$ significantly exceeds the ohmic time $\tau_\sigma^0$, giving ions ample time to recombine. As a result, the heterocharge layer becomes thinner, and the charge cloud does not form. Furthermore, as the applied voltage $\Delta\psi$ increases, the conduction number $W_0$ decreases, which leads to a reduction in charge accumulation on the droplet interface and weakens the electric field force. Therefore, the droplet deformation gradually diminishes progressively as the applied voltage $\Delta\psi$ rises.

\begin{figure}[H]
	\centering
	\includegraphics[width=0.5\textwidth]{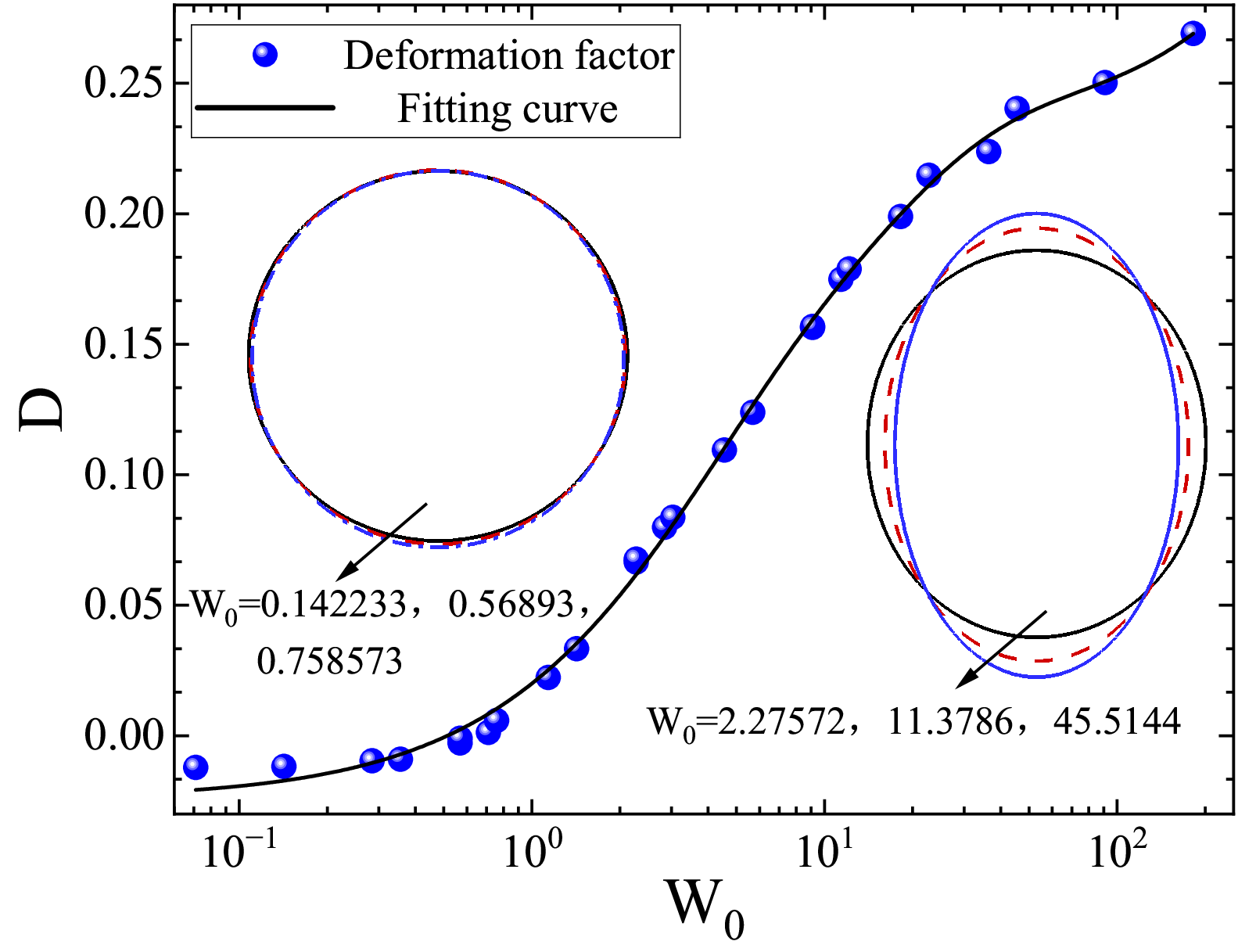}
	\caption{The relationship between conduction number $W_0$ and deformation factor $D$, in which the insets show the variation of interface positions with conduction number $W_0$.}
	\label{fig15}
\end{figure}

Current analysis reveals that the reference length $d$ and the applied voltage $\Delta\psi$ influence droplet deformation by changing the conduction number $W_0$. Building on this, the deformation factor $D$ [see Eq. (\ref{eq63})] is introduced to provide a quantitative measure of the influence of the conduction number $W_0$ on droplet deformation. According to the definition of $D$, the droplet becomes a prolate shape when $D\textgreater 0$, and an oblate shape when $D\textless 0$. Fig. \ref{fig15} depicts the relationship between $W_0$ and $D$, showing that $D$ increases as $W_0$ rises. Specifically, when $W_0\textless 1$, the deformation factor is close to zero, indicating slight droplet deformation. This occurs because a smaller conduction number $W_0$ results in a shorter ion transit time $\tau_K$, which prevents the formation of a distinct electroneutral region within the droplet. Additionally, the uniform charge distribution on the droplet interface creates a uniform electric field force, leading to negligible deformation. When $W_0\textgreater1$, the deformation factor $D$ increases as the conduction number $W_0$ rises. As the ion transit time $\tau_K$ increases, an electroneutral region gradually forms inside the droplet, and the charge accumulates more at the droplet interface. The resulting nonuniform charge distribution on the droplet interface induces an uneven electric field force, which leads to the droplet deformation increasing gradually.

\begin{figure}[H]
	\centering
	\subfigure[]{\label{fig5a}\includegraphics[width=0.3135\textwidth]{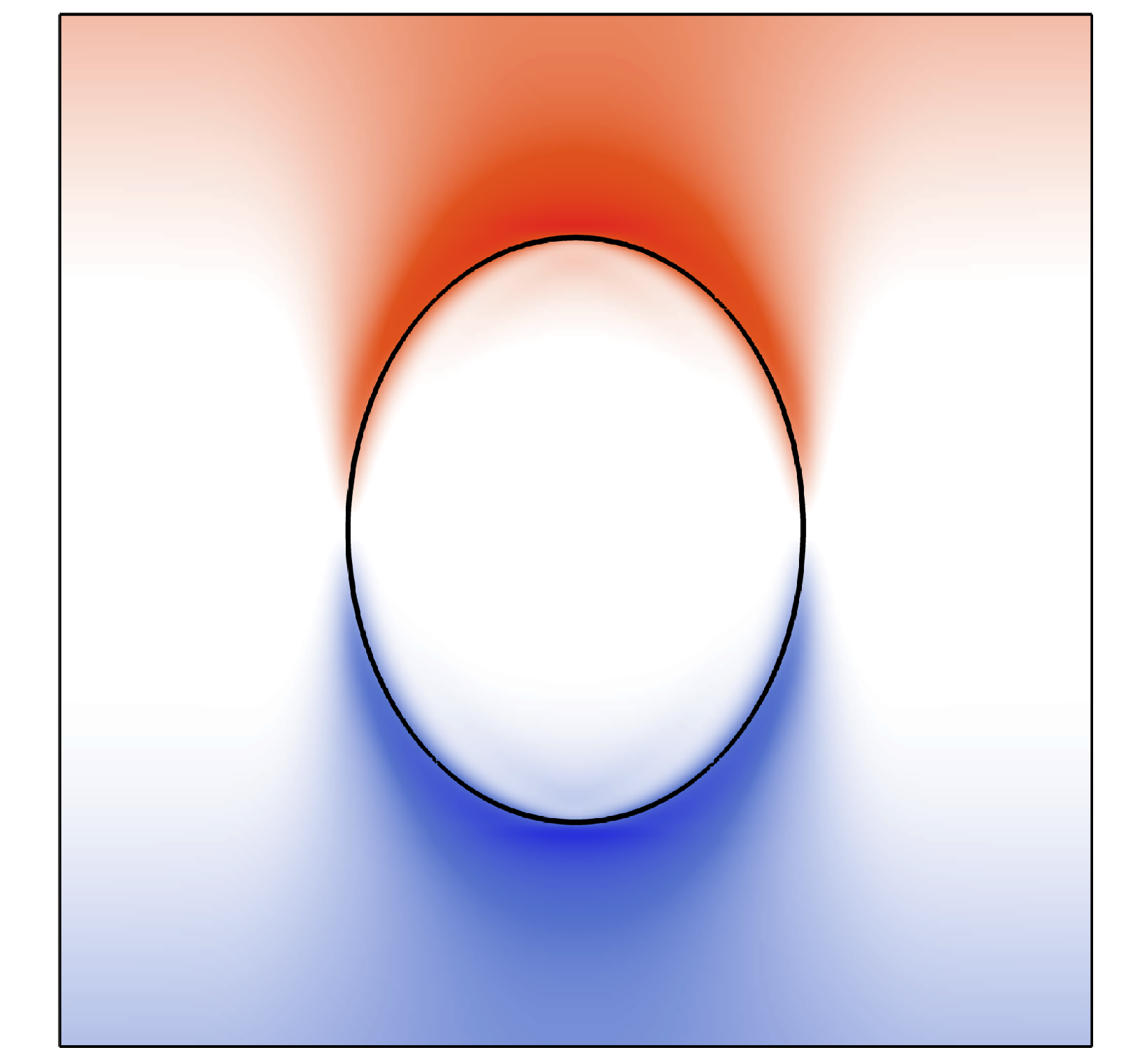}}
	\subfigure[]{\label{fig5b}\includegraphics[width=0.314\textwidth]{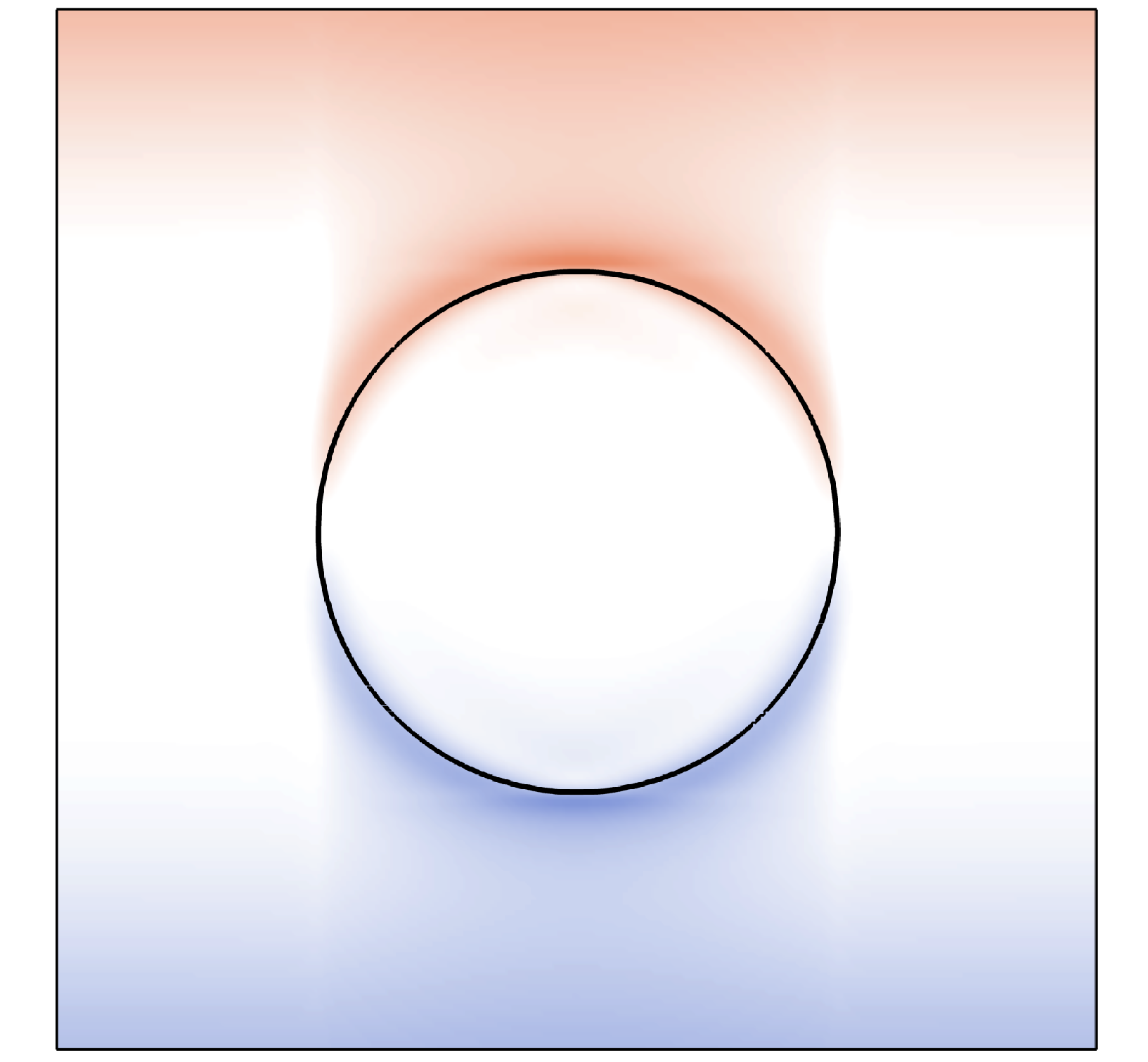}}
	\subfigure[]{\label{fig5c}\includegraphics[width=0.352\textwidth]{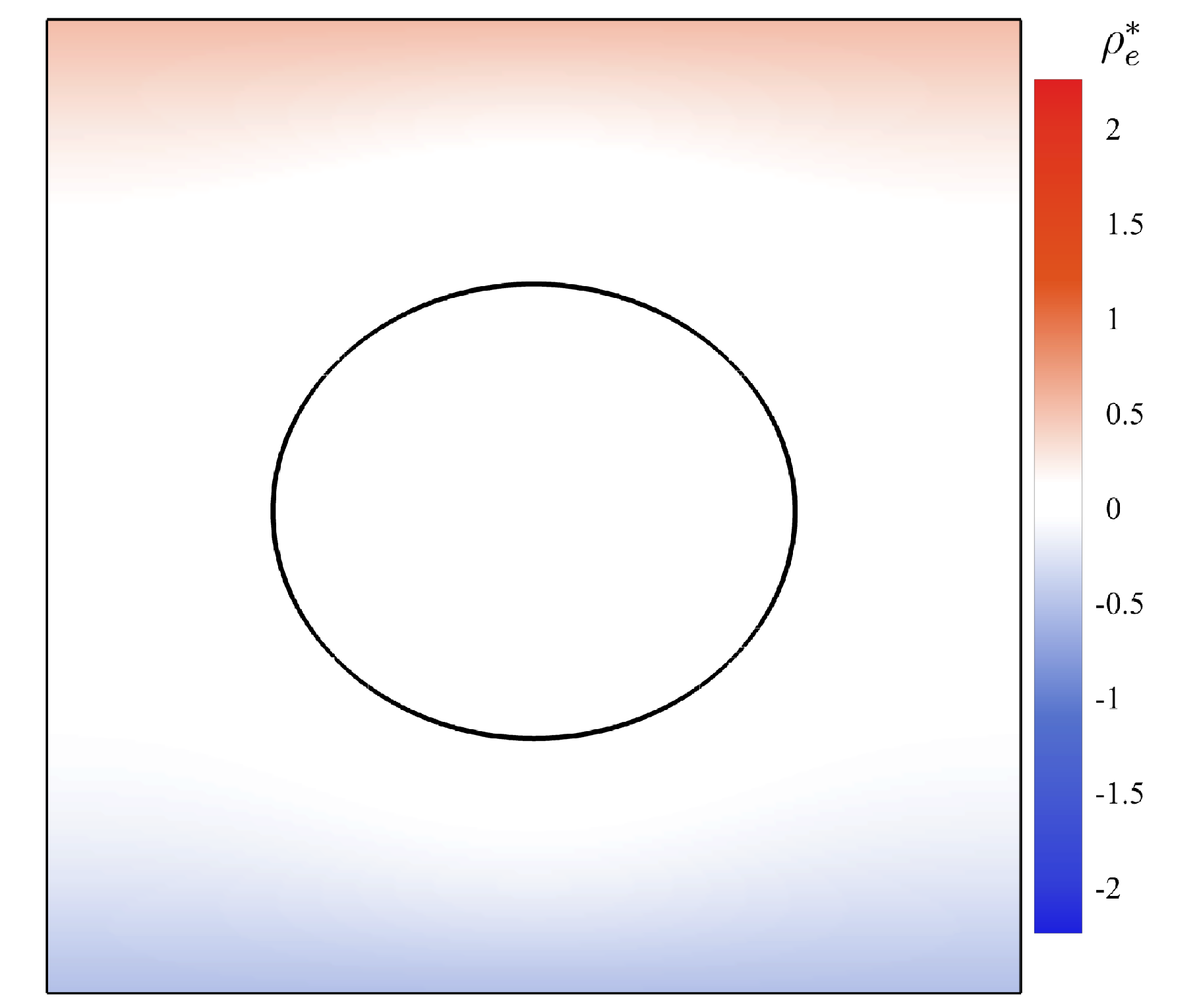}}\\
	\caption{The charge distribution (colored regions) and the droplet profile (black lines) for three different values of the permittivity ratios: (a) $\varepsilon_r=0.5$; (b) $\varepsilon_r=5.0$; (c) $\varepsilon_r=50.0$.}
	\label{fig5}
\end{figure}

\begin{figure}[H]
	\centering
	\subfigure[]{\label{fig6a}\includegraphics[width=0.482\textwidth]{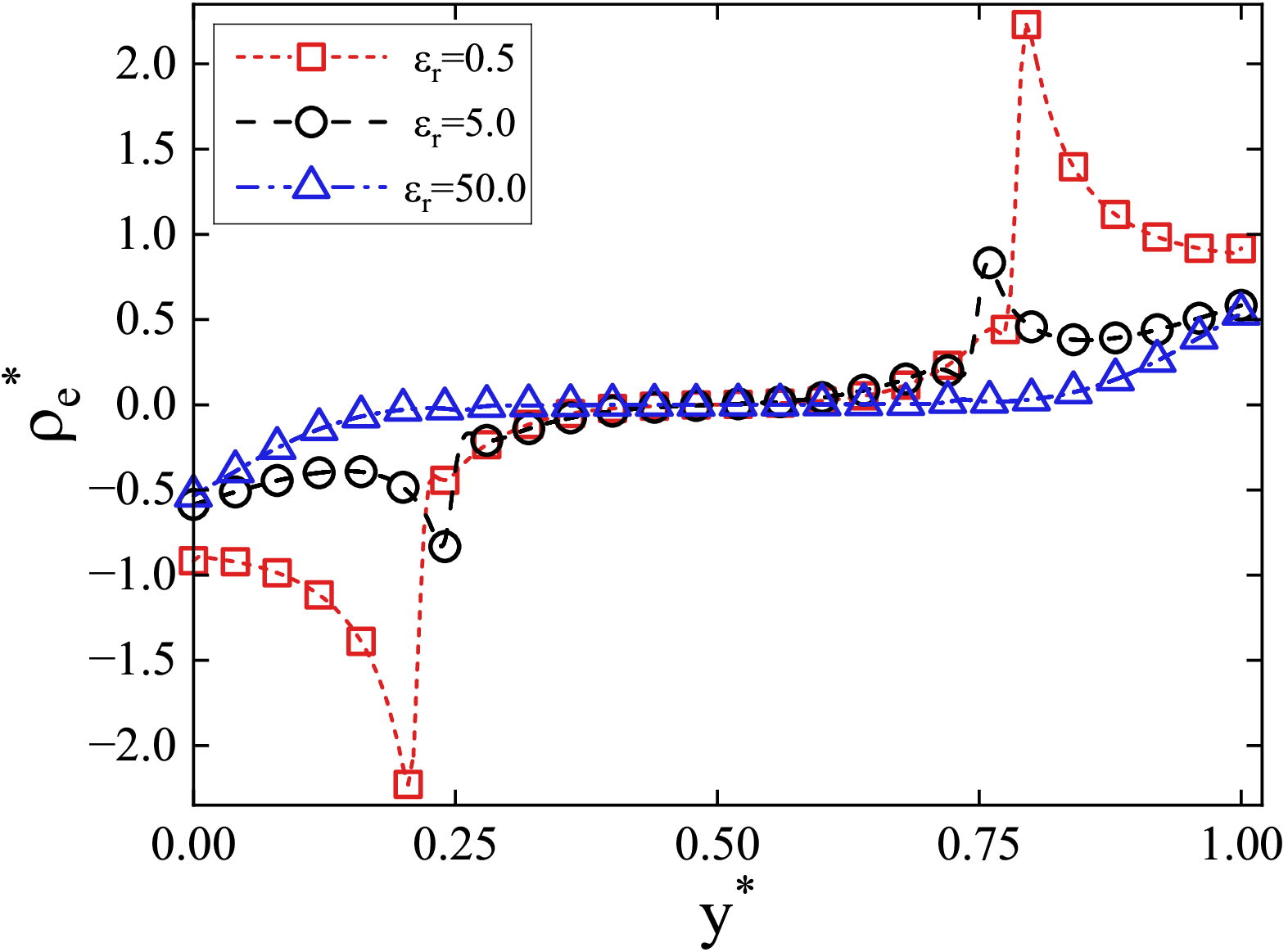}}
	\subfigure[]{\label{fig6c}\includegraphics[width=0.488\textwidth]{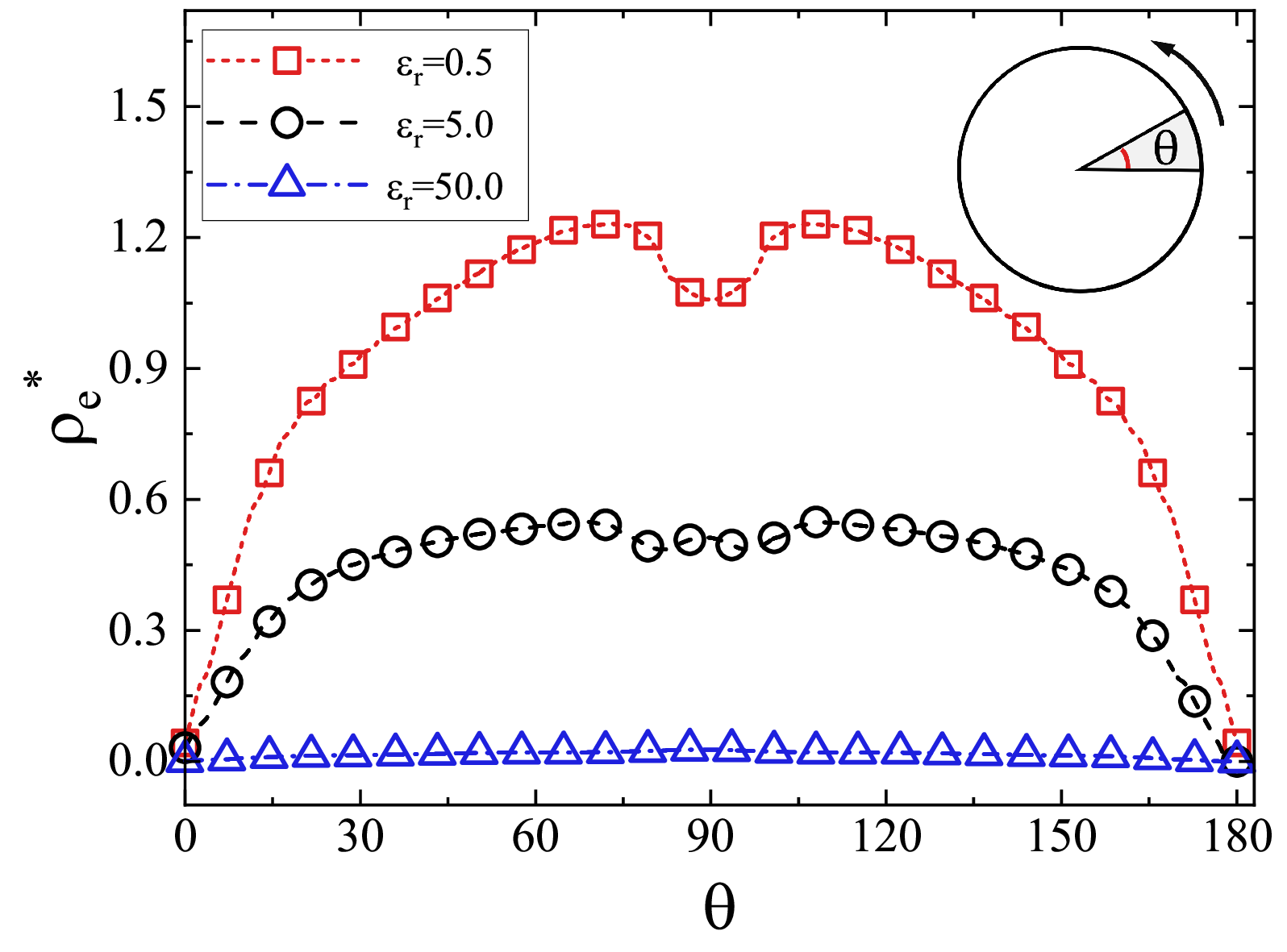}}		\\
	\caption{The distributions of the charge for three different permittivity ratios: (a) $x=0.5d$ of the cavity, (b) the droplet interface, in which $\theta$ represents the angle of the charge distribution at the droplet interface and the inset show the direction of rotation of $\theta$.}
	\label{fig6}
\end{figure}

The above discussion is conducted by fixing the permittivity ratio $\varepsilon_r$ and ionic mobility ratio $\mu_r$. Similar to the droplet deformation within the framework of the leaky dielectrics, it is expected that the permittivity ratio and ionic mobility ratio variations could also affect the shape of the droplet induced by the Onsager-Wien effect. To fully understand droplet deformation in this case, Fig. \ref{fig5} presents the distributions of charge density for three representative permittivity ratios with $d=2.0$ mm and $\Delta\psi=1.0$ KV. It is noted that as $\varepsilon_r=0.5$, the fluid with higher permittivity has stronger polarization under the applied electric field, resulting in a relatively stronger electric field inside the droplet. In this case, cations are attracted to the cathode and primarily accumulate in the northern hemisphere of the droplet and near the cathode, with a small amount distributed in the southern hemisphere and near the anode, and the anions show the opposite distribution. As a result, negative charges predominantly accumulate in the southern hemisphere of the droplet, while positive charges concentrate in the northern hemisphere (see Fig. \ref{fig5a} or Fig. \ref{fig6a}). This charge distribution further generates a greater electric field force in the vertical direction, leading to the droplet being a prolate shape. As the permittivity ratio $\varepsilon_r$ increases to 5.0, the electric field within the droplet experiences a gradual weakening. Consequently, the charge difference between the electrode and the droplet interface becomes relatively small, resulting in a reduction of charge difference along the droplet interface (see Fig. \ref{fig5b} or Fig. \ref{fig6c}). This leads to a more balanced electric force acting on the droplet, and the droplet does not experience significant deformation. When the permittivity ratio $\varepsilon_r$ further increases to 50.0, the electric field inside the droplet weakens further, and the charge accumulated at the droplet interface significantly decreases. Therefore, most of the charge concentrates near the electrodes, forming a heterocharge layer of opposite charges (see Fig. \ref{fig5c} or Fig. \ref{fig6a}). Under these conditions, the droplet experiences a larger horizontal electric force, causing the droplet to be an oblate shape.

\begin{figure}[H]
	\centering
	\subfigure[]{\label{fig8a}\includegraphics[width=0.313\textwidth]{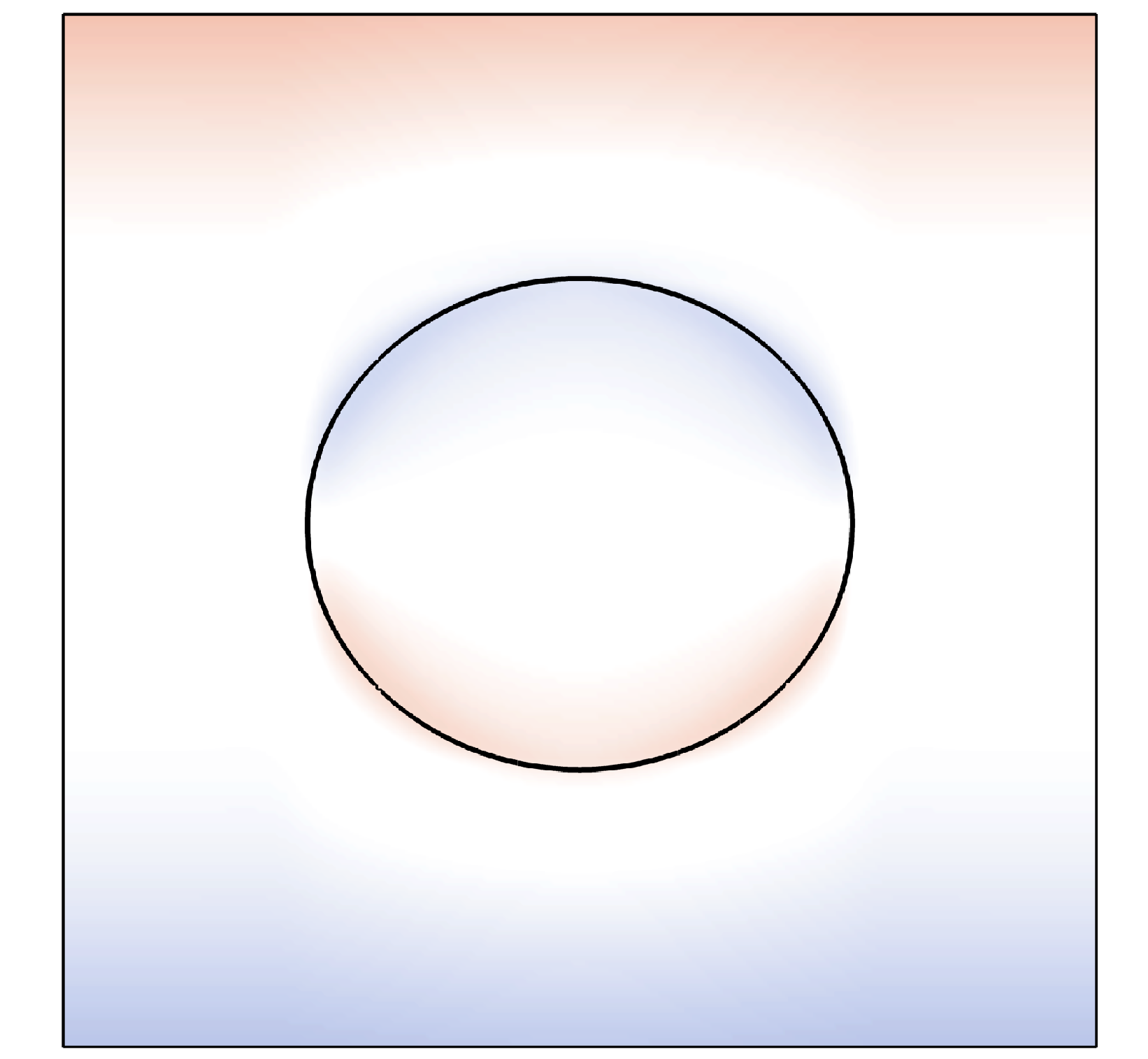}}		\subfigure[]{\label{fig8b}\includegraphics[width=0.3135\textwidth]{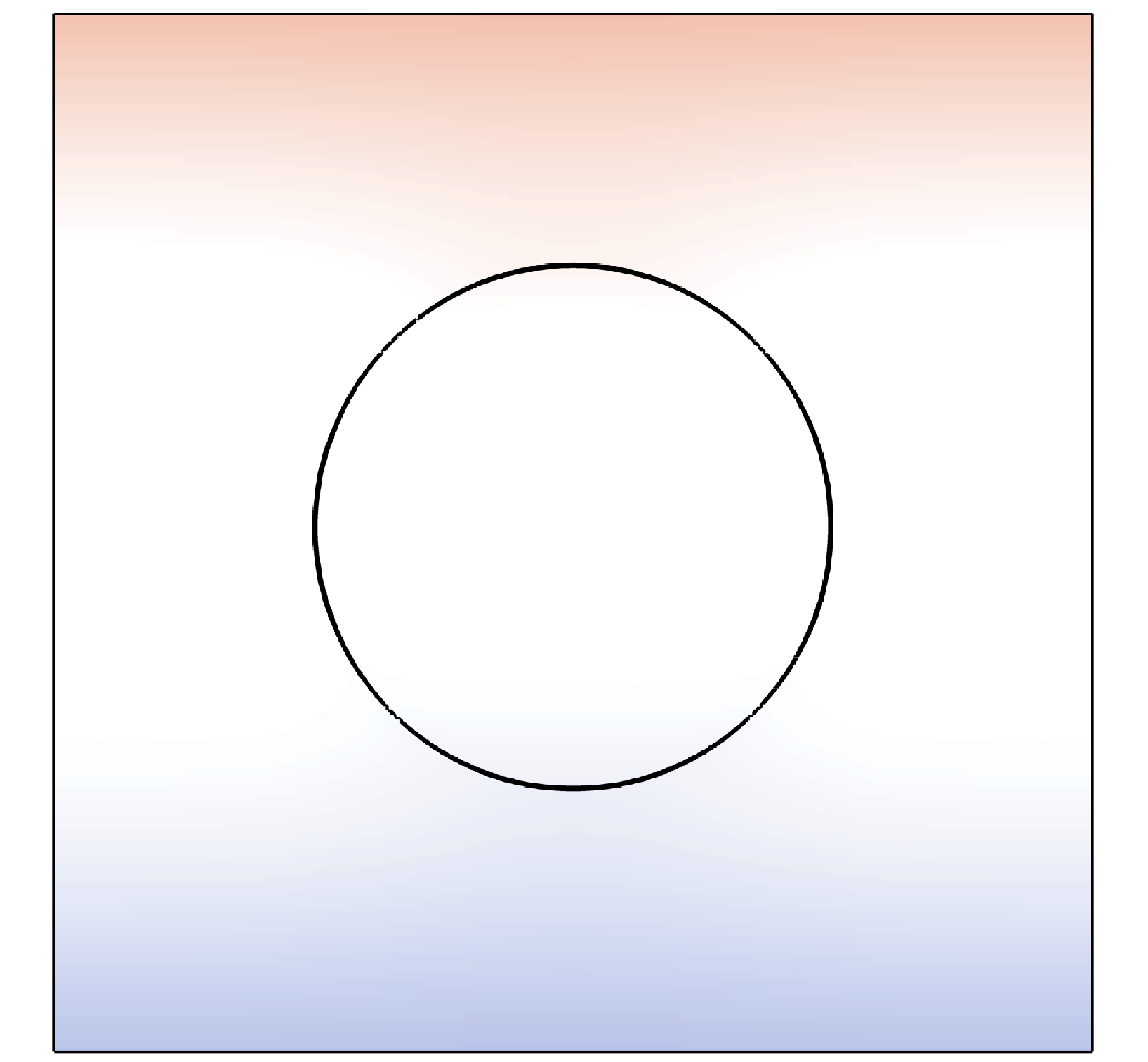}}
	\subfigure[]{\label{fig8c}\includegraphics[width=0.352\textwidth]{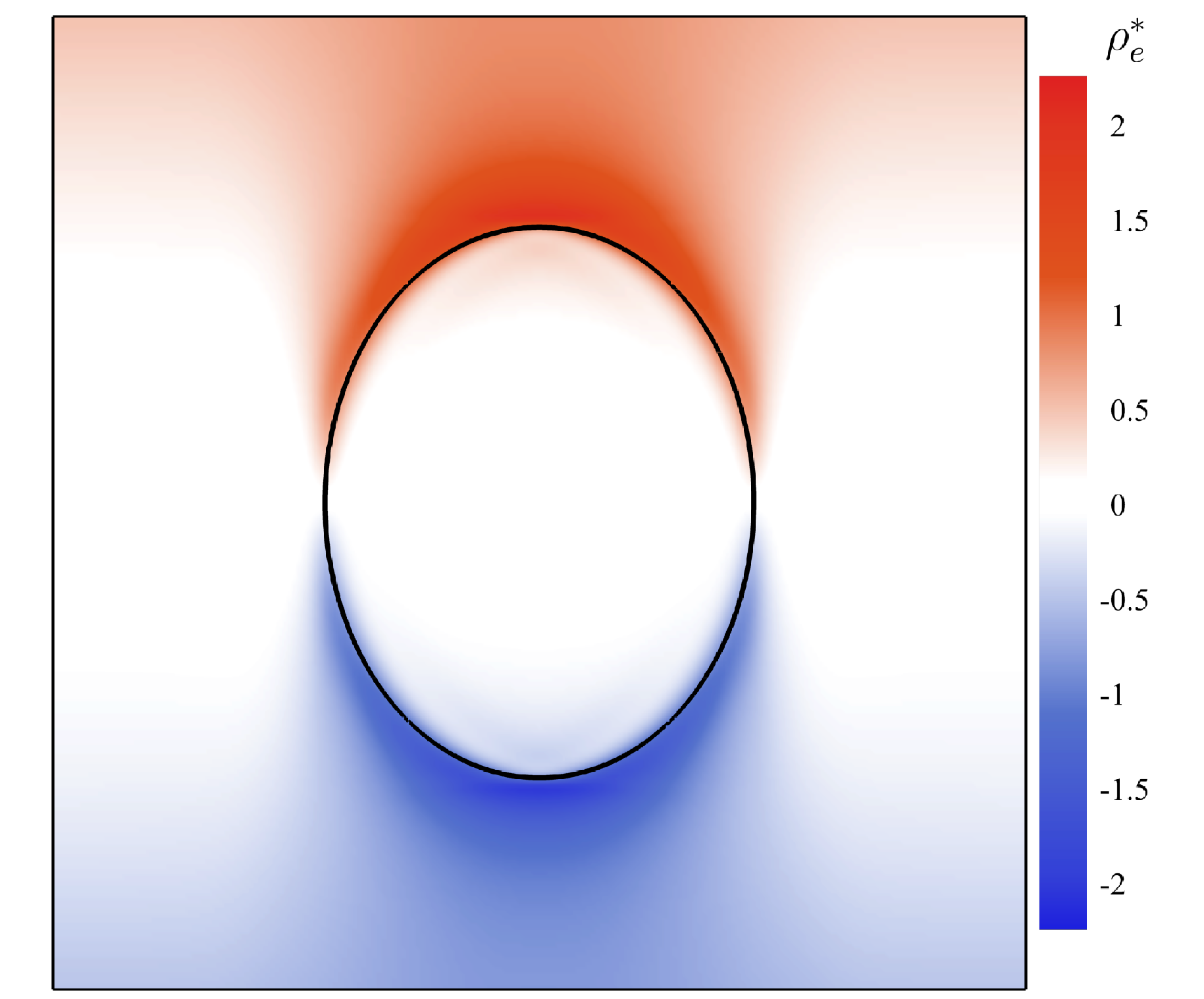}}\\
	\caption{The charge distribution (colored regions) and the droplet profile (black lines) for three different values of the ionic mobility ratios: (a) $\mu_r=0.5$; (b) $\mu_r=1.0$; (c) $\mu_r=5.0$.}
	\label{fig8}
\end{figure}

\begin{figure}[H]
	\centering
	\subfigure[]{\label{fig9a}\includegraphics[width=0.486\textwidth]{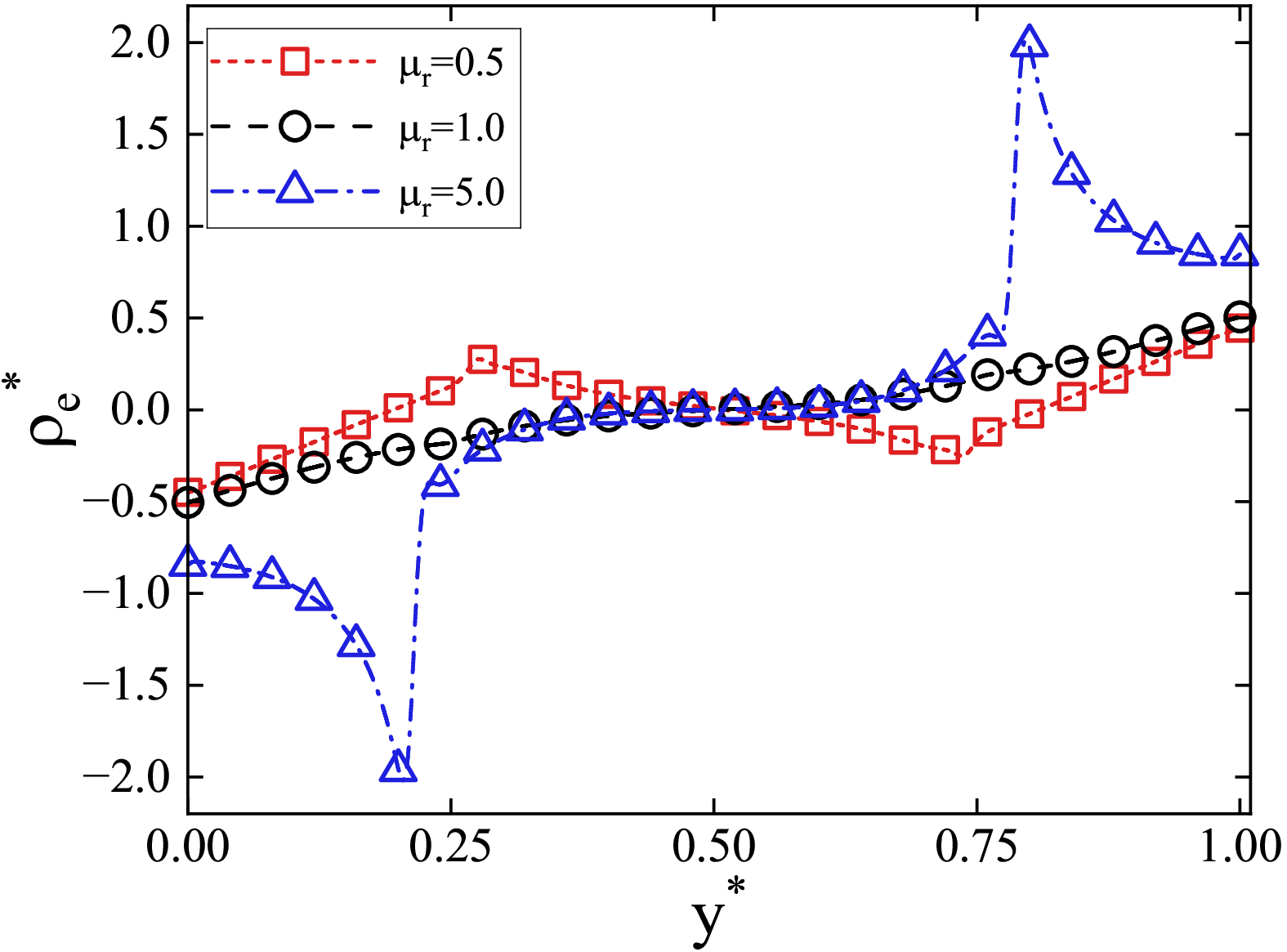}}	
	\subfigure[]{\label{fig9b}\includegraphics[width=0.49\textwidth]{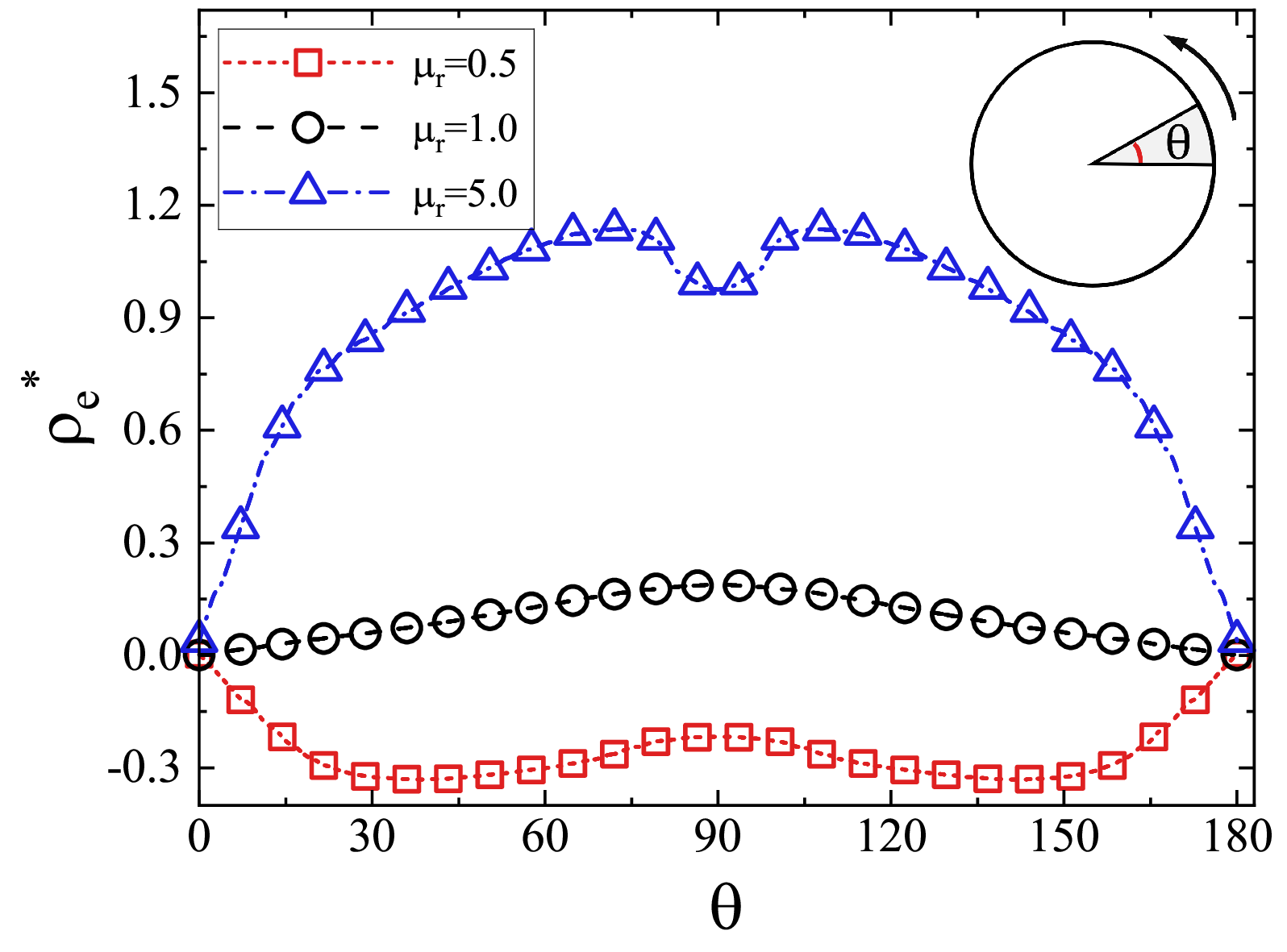}}	
	\caption{The charge distribution for three different ionic mobility ratios: (a) $x=0.5d$ of the cavity, (b) the droplet interface, in which $\theta$ represents the angle of the charge distribution at the droplet interface and the inset show the direction of rotation of $\theta$.}
	\label{fig9}
\end{figure}

After analyzing the effects of the permittivity ratio, we further investigate the effect of ionic mobility ratio on droplet deformation, and the results are depicted in Fig. \ref{fig8} and Fig. \ref{fig9}. It is observed that as $\mu_r = 0.5$, the lower ionic mobility within the droplet limits the migration of ions toward the electrodes. As a result, most cations accumulate in the southern hemisphere of the droplet and near the cathode, and a small number of cations are distributed near the anode, while the anions exhibit the opposite distribution. This leads to the formation of a heterocharge layer of opposite charges near the electrodes, where positive charges accumulate in the southern hemisphere of the droplet, while negative charges concentrate in the northern hemisphere (see Fig. \ref{fig8a} or Fig. \ref{fig9a}). The resulting charge distribution generates an electric force that compresses the droplet along the vertical axis, which causes the droplet to adopt an oblate shape. As ionic mobility ratio increases to 1.0, cations primarily accumulate near the cathode and the droplet interface, with fewer present near the anode, while anions exhibit the opposite distribution. At this stage, most charges are concentrated near the electrodes, and the concentrations of cations and anions near the droplet interface become nearly equal. Consequently, the droplet interface shows almost no charge accumulation (see Fig. \ref{fig8b} or Fig. \ref{fig9b}). Under these conditions, the electric field force acting on the droplet weakens, leading to the droplet having negligible deformation (see Fig. \ref{fig8b}). When ionic mobility ratio further increased to 5.0, charge accumulation at the droplet interface becomes more pronounced (see Fig. \ref{fig8c}). This induces the droplet to experience a stronger electric force in the vertical direction, causing the droplet to adopt a prolate shape.

\begin{figure}[H]
	\centering
	\subfigure[]{\label{fig17a}\includegraphics[width=0.483\textwidth]{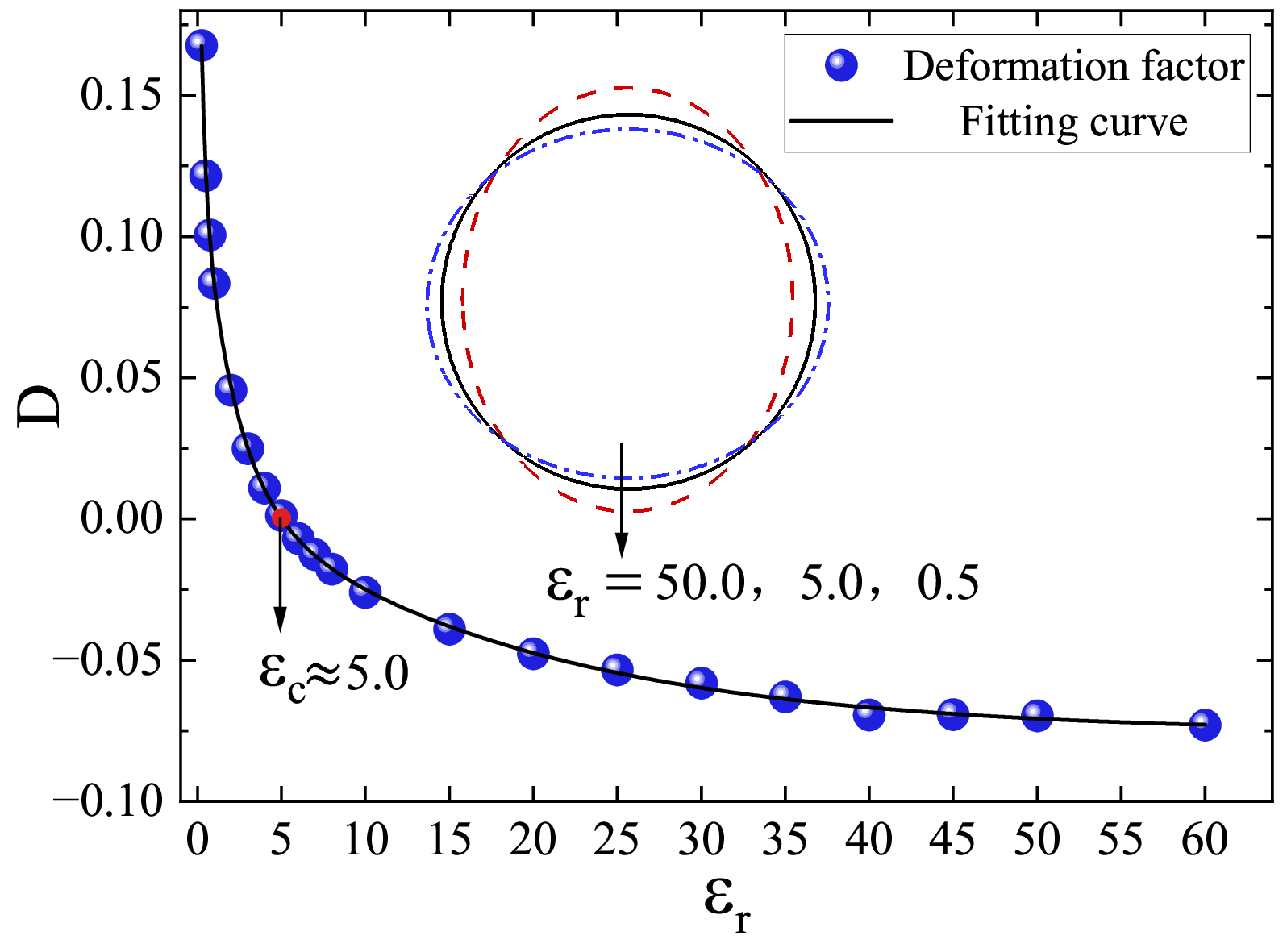}}	
	\subfigure[]{\label{fig17b}\includegraphics[width=0.481\textwidth]{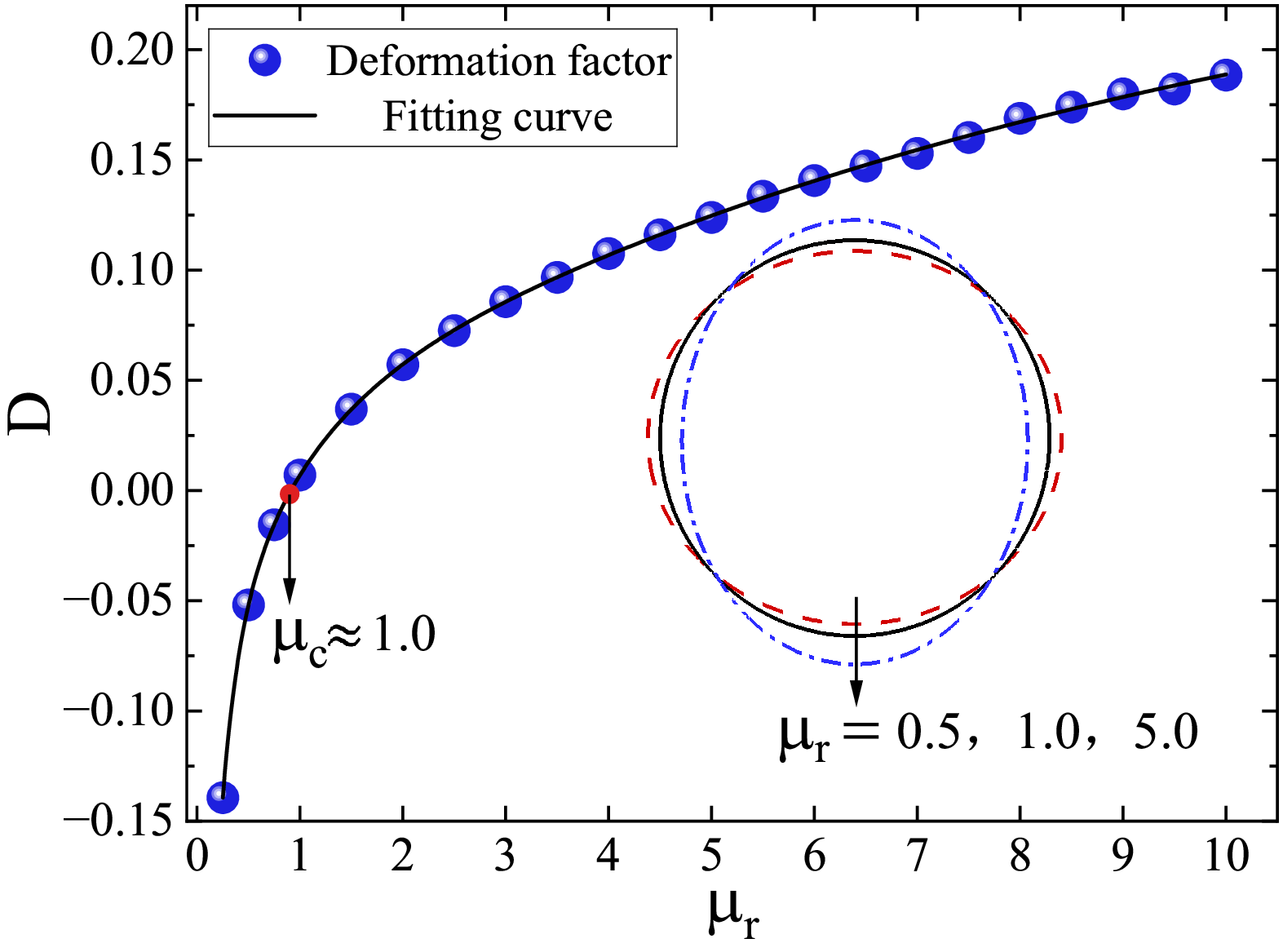}}	
	\caption{The relationship of the permittivity ratio $\varepsilon_r$ (a) and the ionic mobility ratio $\mu_r$ (b) with the deformation factor $D$, in which the inset show the variation of interface positions with the permittivity ratio $\varepsilon_r$ (a) or the ionic mobility ratio $\mu_r$ (b). $\varepsilon_c$ and $\mu_c$ represent the critical values for the permittivity ratio $\varepsilon_r$ and the ionic mobility ratio $\mu_r$, respectively.}
	\label{fig17}
\end{figure}

Finally, to understand the development of the deformation factor $D$ induced by the Onsager-Wien effect, we also carry out full simulations with different permittivity ratios and ionic mobility ratios and present the numerical results in Fig. \ref{fig17}. As seen from this figure, the deformation factor decreases with an increase in the permittivity ratio, but it exhibits a positive correlation with the ionic mobility ratio. Another point that needs to be emphasized is that the droplet always adopts a prolate/oblate shape when $\varepsilon_r\textless\varepsilon_c$/$\mu_r\textless\mu_c$ (here, $\varepsilon_c$ and $\mu_c$ are the critical permittivity ratio and ionic mobility ratio, at which the droplet shares a similar shape to its initial state). However, when $\varepsilon_r\textgreater\varepsilon_c$/$\mu_r\textgreater\mu_c$, the droplet shares an oblate/prolate shape.

\section{Conclusions}\label{section7}
In this paper, we propose an LB model of two-phase EHD flows induced by the Onsager-Wien effect. The model adopts five distribution functions to solve the conservative Allen-Canh equation for capturing the phase interface, the electric field equations, the Nernst-Planck equations for ionic transport, and the Navier-Stokes equations describing the fluid flow. To validate the feasibility of this model in the EHD conduction problem induced by the Onsager-Wien effect, we simulate a one-dimensional permeable electrode problem submerged in a dielectric fluid. Subsequently, we simulate the deformation of a leaky dielectric droplet to assess further its effectiveness in solving the two-phase EHD problem. The numerical results demonstrate that the solutions to both simulated problems agree well with analytical or existing numerical data, which generates satisfactory results.

We further use the proposed LB model to analyze the droplet deformation induced by the Onsager-Wien effect. This analysis includes a detailed discussion of the effects of the reference length \( d \), the applied voltage \( \Delta \psi \), the permittivity ratio \( \varepsilon_r \), and the ionic mobility ratio \( \mu_r \). Unlike droplet deformation in the leaky dielectric model, the EHD conduction mechanism induced by the Onsager-Wien effect clearly shows heterocharge layers near the electrodes. Additionally, the numerical results show that both the reference length $d$ and the applied voltage $\Delta\psi$ further influence the droplet deformation and charge distribution in the presence of the conduction number $W_0$. Specifically, with the conduction number $W_0$ increasing, droplet deformation becomes more significant, while both the thickness of the heterocharge layer near the electrode and the concentration of the charge cloud decrease. Furthermore, critical values $\varepsilon_c$ and $\mu_c$ exist for droplet deformation,  corresponding to the permittivity ratio $\varepsilon_r$ and the ionic mobility ratio $\mu_r$. When the permittivity ratio $\varepsilon_r$ (or the ionic mobility ratio $\mu_r$) exceeds the critical value, the droplet adopts an oblate (or prolate) shape. Conversely, when the ionic mobility ratio $\mu_r$ (or the permittivity ratio $\varepsilon_r$) is below the critical value, the droplet transitions a prolate (or oblate) shape. While near the critical value, the droplet experiences minimal deformation. Finally, since the current study does not address wetting behavior, we plan to examine the influence of two-phase flows, driven by the Onsager-Wien effect, on wetting behavior in future work to further enrich the study.

\section*{Acknowledgments}
This work is financially supported by the National Natural Science Foundation of China (Grant No. 12472297).

\end{document}